\documentclass[10pt,preprint]{aastex}
\input epsf.tex
\begin{document}
%% Prints a large "DRAFT" diagonally across each page
%% Does not show up in TeXview
% \typeout{Prints "DRAFT" on each page; does not show in TeXView}
% \special{!userdict begin /bop-hook{gsave 200 30 translate
% 65 rotate /Times-Roman findfont 216 scalefont setfont
% 0 0 moveto 0.95 setgray (DRAFT) show grestore}def end}
\title{Migration then assembly: Formation of Neptune mass planets inside 1~AU} 
\author{Brad M. S. Hansen\altaffilmark{1} \& Norm Murray\altaffilmark{2,3}
}
\altaffiltext{1}{Department of Physics \& Astronomy and Institute of Geophysics \& Planetary Physics, University of California Los Angeles, Los Angeles, CA 90095, hansen@astro.ucla.edu}
\altaffiltext{2}{Canadian Institute for Theoretical Astrophysics, 60 St George St, Toronto, Ontario, Canada}
\altaffiltext{3}{Canada Research Chair in Astrophysics}

%\slugcomment{\it 
%}

\lefthead{Hansen \& Murray}
\righthead{Migrate, then Assemble}

\begin{abstract}
We demonstrate that the observed distribution of `Hot Neptune'/`Super-Earth' systems is well reproduced by a model in which
planet assembly occurs in situ, with no significant migration post-assembly. This is achieved only if the amount of mass in rocky material
is $\sim 50$--$100 M_{\oplus}$ interior to 1~AU. Such a reservoir of material implies that significant radial migration of
solid material takes place, and that it occur before the stage of final planet assembly.
 The model not only reproduces the general distribution of mass versus period,
but also the detailed statistics of multiple planet systems in the sample.

We furthermore demonstrate that cores of this size are also likely to meet the criterion to gravitationally capture
gas from the nebula, although accretion is rapidly limited by the opening of gaps in the gas disk. If the mass growth is
limited by this tidal truncation, then the scenario sketched here naturally produces Neptune-mass objects with substantial components
of both rock and gas, as is observed.

The quantitative expectations of this scenario are that most planets in the `Hot Neptune/Super-Earth' class inhabit multiple-planet
systems, with characteristic orbital spacings. The model also provides a natural division into gas-rich (Hot Neptune) and gas-poor (Super-Earth) classes at fixed period. The dividing
mass ranges from $\sim 3 M_{\oplus}$ at 10~day orbital periods to $\sim 10 M_{\oplus}$ at 100~day orbital periods. For orbital periods
$< 10$~days, the division is less clear because a gas atmosphere may be significantly eroded by stellar radiation.

\end{abstract}


\section{Introduction}

The discovery of the planet around 51 Pegasi (Mayor \& Queloz 1995) has spurred an avalanche of observations over the
last 15 years, leading to a census of almost 700 confirmed planets to date. Within this data we can find a variety
of exotic and unexpected systems, such as planets with short orbital periods, extreme orbital eccentricities, or systems
with strong resonant interactions. These observations have also inspired a wealth of theoretical work, on both the origin
and structure of such planets. Furthermore, planet searches have continued to push the observational thresholds towards 
lower masses, with current observations now probing below $5 M_{\oplus}$ 
(Mayor et al. 2009a,b; Leger et al. 2009; Howard et al. 2011; Batalha et al. 2011), 
with the hope of eventually finding planets that may convincingly be argued to be hospitable to life.

Despite the wealth of data and interpretation now available, it is still difficult to draw together the different elements
into an overall synthesis, in large part because of the heterogeneous nature of the observations. However, in the next
few years a variety of observations offer the possibility to provide new constraints, which may allow us to frame a
rigorous context for planet formation and evolution.  Recently, Howard et al. (2010) and Mayor et al. (2011) published the results of volume 
limited radial velocity surveys of nearby  sun-like stars. The importance of these studies is that one can now begin to compare 
the relative frequencies of different types of planets, illuminating which are the generic product of the planet formation process, and which are the exotica.
The results suggest that lower mass planets ($<100 M_{\oplus}$) are even more common than gas giants at short periods, 
and that they appear to form a distinct population for orbital periods $< 50$~days. Particularly striking is the fact 
that the low-mass planet distribution appears to show an upper edge in a plot of mass versus orbital period;
%(see Figure~\ref{Ma1}); 
planets are found with masses between the completeness limit of the survey and an upper limit ranging between $7M_\oplus$ at 
$P\sim2$~day and $20M_\oplus$ at $P\sim50$~days. This is in contrast to the results of the predominant planet population synthesis models 
(e.g. Ida \& Lin 2008; Mordasini, Alibert \& Benz 2009), which predict that this mass range ($10-50 M_{\oplus}$) should
be liberally populated with 
  partially-accreted giant planet cores (``hot Neptunes'') that have migrated in from larger radii.

In addition to the radial velocity surveys, the Kepler satellite has begun to release the contents of its large-scale transit
catalogue (Borucki et al. 2011, Lissauer et al. 2011a). The majority of the present sample are still unconfirmed `candidates', 
but a variety of tests suggest that the true planetary yield from this sample is expected to be high (Borucki et al. 2011). 
The Kepler sample once again has the advantage of being drawn from a relatively homogeneous program, and the distribution of 
planetary radii supports the aforementioned claim that the dominant species of short period planets corresponds to radii and 
masses below that expected from gas giant planets (Howard et al. 2011; Wolfgang \& Laughlin 2011; Youdin 2011).

These new surveys highlight  the importance of short-period low-mass planets, as the low-mass planets now appear to outnumber 
their more massive brethren. The origins of such 'Hot Neptunes' and `super-Earths' is still poorly understood. Most models 
of their formation (Alibert et al. 2006; Terquem \& Papaloizou 2007; Raymond, Barnes \& Mandell 2008; Mordasini et al. 2009;  Ida \& Lin 2010; McNeil \& Nelson 2010) 
adopt the position that they form at larger radii, in a manner related (if not similar) to the planets Uranus and Neptune in 
our own solar system, and then migrate inwards, to be trapped closer to the star. However, quantitative comparisons with the 
properties of observed systems remain unsatisfactory (McNeil \& Nelson 2010).

Ultimately, many of the problems with these calculations stem from the properties of the migration models assumed. Although the
notion of gaseous disk migration has a long, physically motivated provenance (e.g. Goldreich \& Tremaine 1980; Ward 1997), the nature
of the migration and when it occurs is still a subject of significant uncertainty (see Kley \& Nelson 2012 for a recent review).
 The principal reason for the revival of interest in migration models was
the notion that hot Jupiters were emplaced in this fashion (e.g. Lin, Bodenheimer \& Richardson 1996).
 However, it is becoming clear that these systems represent a small minority of all planetary systems (e.g. Howard et al. 2010; Mayor et al. 2011)
and may therefore reflect rather special circumstances. Indeed, there is a growing
 accumulation 
of recent results demonstrating the misalignment of stellar spin and planetary orbits in many hot Jupiter systems
(e.g. Winn et al. 2011a and references therein). The simplest interpretation is that at least some hot Jupiters did not migrate through a gas disk. 
Furthermore, it is no longer clear that migration implies a monotonic inward motion (e.g. Paardekooper \& Mellema 2006; Oishi, Mac Low \& Menou 2007;
Adams \& Bloch 2009). With 
the large numbers of $5-20M_\oplus$ planets now seen at short periods, where the simplest migration models suggest they should not be, indicates 
that large radial movement need not be ubiquitous in planetary systems.

 Motivated
by this renewed skepticism, we have chosen to revisit the question of in situ accumulation of rocky planets on small scales, and
to explore the consequences of planet assembly without significant late-time gaseous migration. However, we will consider the possibility that
significant radial migration is possible for smaller bodies, before they accumulate into planets, and so we will allow for disks with
a wide variety of masses, discussed in \S~\ref{Inventory}.
In \S~\ref{Accum} we then present the results of a series of traditional in situ planetary accumulation models on scales of 0.05--1~AU, from the resulting disks.
In \S~\ref{Accretion} we investigate the consequences of gas accretion onto the massive rocky cores that form.
We address the role of Jupiter mass planets in \S~\ref{Hotties} and evaluate the possible role of photoevaporation in \S~\ref{Evap}.

\section{How much rocky material is available for assembly?}
\label{Inventory}

The prototypical initial condition for assembly of a planetary system is the `Minimum-Mass Solar Nebula' (Weidenschilling 1977a; Hayashi 1981), based
on an estimate of the material required to form our own planetary system. The required surface density of solids (based here on the Hayashi estimate) is
\begin{equation}
 \Sigma_Z = 7  \left( \frac{a}{1 AU} \right)^{-3/2} {\rm g}\, {\rm cm}^{-2},
\end{equation}
with the gas surface density $\sim 240$ times larger.
This applies interior to the `snow line' (taken to be at $a=3$~AU). Exterior to the snow line the available material in solids is a factor of four higher, reflecting
the fact that more volatile elements are able to condense and form solids as the nebular temperature drops. This enhancement brings
the solid/gas ratio closer to the estimated metallicity of the sun, which presumably reflects the original composition of the nebular gas.
If we integrate this surface density we infer a mass in solids of 3.3$M_{\oplus}$ interior to 1~AU, $5.7 M_{\oplus}$ inside the snow line, and
25.5 $M_{\oplus}$ within  10~AU.
This inventory is the minimum amount of mass required to match the solar system planets, but factors of two or three more are not unreasonable, if we assume
a certain inefficiency in the process of converting nebular material into planets. In addition, many planet hosting stars are known to show super-solar
atmospheric abundances (Gonzalez 1998; Santos, Israelian \& Mayor 2001; Fischer \& Valenti 2005; Udry \& Santos 2007), often up to a factor of two. As such, we can reasonably argue for an increase in the above inventory of solids by a factor of
five. Thus, the simplest extension of the MMSN mass accounting  could plausibly provide 17$M_{\oplus}$ inside 1~AU, 29$M_{\oplus}$ inside the snow line, and 128$M_{\oplus}$ inside 10~AU.

The above estimates are still very much in the spirit of the original MMSN model, in which solid material condenses out of a gaseous nebula and
accumulates in situ to form planets from the local reservoir of material. However, the modern view of nebular and planetary evolution is considerably
more dynamic. Recent models have focused largely on the migration of planet-sized bodies, but it has been known for a long time that the dust and small
rocks from which planets form are also potentially subject to 
 significant radial migration (Weidenschilling 1977b; Nakagawa, Sekiya \& Hayashi 1986; Youdin \& Shu 2002; Bai \& Stone 2010). If a significant fraction
of the solid material can migrate interior to 1~AU while it is still in the form of small bodies, then the potential reservoir of material for the
 accumulation of terrestrial-class planets could be considerably larger. We will consider mass reservoirs up to 100 $M_{\oplus}$, corresponding to
a radial concentration of mass by factors of at least 10. The nature of the processes that produce such radial concentrations is still a subject of active
research, and it is not guaranteed that the solids will preserve the same radial profile as the original gas disk.
Thus we will consider solid surface density profiles with a range of slopes.

\section{Rocky Planet Accumulation Results}
\label{Accum}

There is a long history of studies of terrestrial planet accumulation in our own solar system. We adopt the oligarchic model of
Kokubo \& Ida (1998) as our starting point, which provides an analytic prescription for a protoplanetary system consisting of a
 sequence of annuli, each dominated by a single planetary embryo.
  Simulations of the last stages of gravitational interaction and accumulation of embryos into planets in this model, 
using N-body techniques (Chambers
\& Wetherill 1998, Agnor, Canup \& Levison 1999; Chambers 2001; Raymond, Quinn \& Lunine 2006), shows that configurations of three or four planets emerge naturally from
a minimum mass solar nebula disk, although the low masses of Mercury and Mars in
  the Solar System may require a more localised initial distribution of mass (Hansen 2009). The production of 
multi-planet systems is fairly robust with respect to varying the amount of mass and the disk density profile
(Raymond, Quinn \& Lunine 2005; Kokubo, Kominami \& Ida 2006), although the masses of the planets tend to scale with the reservoir mass.

Prior simulations of this process have generally restricted themselves to distances $> 0.5$~AU because of the lack of any
closer planet in the Solar system (and shorter timesteps add unecessarily to the computational burden). This is 
also often motivated on the basis of arguments regarding the chemistry of volatiles and a presumed reduction in the
mass fraction that condenses out of the nebula. However, our knowledge of nebula dynamics and chemistry is sufficiently
poorly constrained that it is prudent to examine the dynamical consequences of different disk masses regardless, and
argue about the underpinnings later. Thus, in the following, we will examine the assembly of terrestrial planet-class
systems from power law surface density distributions
of solid material extending down to 0.05~AU and up to 1~AU.

\subsection{In situ accumulation of close-in terrestrial planets}

We use the N-body code Mercury (Chambers 1999) to simulate the final stages of accumulation of planetary
embryos, born of the oligarchic model, into planets around a $1 M_{\odot}$ star. We have performed a series of simulations of disks that extend from 0.05 to 1~AU, while
varying the total mass and surface density profile, to assess whether such an initial distribution can produce the planetary systems
observed in a homogenously selected sample, such as that by Howard et al. (2010). In order to resolve the orbital dynamics at the inner edge of the disk, we use timesteps of 12~hours! Fortunately,
most of the interesting dynamics in these collisional accumulation simulations occurs within a fixed number of orbits, so that the compact
systems simulated here evolve very rapidly in absolute terms. We evolve our systems to 10~Myr, which is sufficient to produce dynamically stable
final results. We have verified this by evolution of selected systems to 100~Myr and 1~Gyr, and have observed little significant change.
Results of these and subsequent simulations are collected in the Table~\ref{SimParam}.

We consider three surface density profiles, all of the power law form $\Sigma \propto a^{-\alpha}$, and
examine models with $\alpha=0, 3/2$ and $5/2$. This encompasses a wide range of behaviour and
is taken to represent the range of uncertainty in the initial conditions, given the possibility
that this might be sculpted by radial migration of solids from large scale to small. With each choice of $\alpha$, we
normalise the surface density to have a particular mass of solids interior to 1~AU. We choose
three different representative normalisations, of 25, 50 and 100~$M_{\oplus}$. 
%We also investigate a fourth
%class of model, the gravitational instability model, as discussed in appendix~\ref{GI}.

In each case, we start with the surface density profile consisting of a series of protoplanetary embryos
realised according to the oligarchic distribution described in equations~17 \& 18 of Kokubo \& Ida (2002),
spaced out between 0.05~AU to 1~AU. 
For the
disk masses considered here, which are larger than usually assumed, the mass of the initial oligarchs
can sometimes reach several $M_{\oplus}$, and one might be concerned that such bodies are massive
enough that their mutual perturbations would invalidate the assumption of mutually
independent radial annuli. In appendix~\ref{Democratic}, 
 we have also investigated models in which we restrict the
mass of oligarchs to be less than $1 M_{\oplus}$, and we
 found no meaningful difference in the eventual outcomes.

%Figure~\ref{Ma1} shows the population results for each disk profile (summed over 3 realisations each) and for each total mass normalisation. 
%Also shown are shaded regions indicating the approximate values of the planets tabulated by Howard et al. Visual inspection demonstrates that the 
%$\alpha=3/2$ models with total masses in the range 50-100$M_{\oplus}$ yield the best qualitative fit to the observed population. 
%The $\alpha=5/2$ models are too steep, so that they populate the observed gap when they match the planets at larger orbital periods 
%($\sim 100$ days). The $\alpha=0$ models do not produce the observed short period planets, with nearly all the mass concentrated at larger radii. 
%The Q=1.5 gas disk models (described in appendix A) are also a very poor fit to the lower mass distribution, but do produce models that are 
%intriguingly close to the {\em upper} mass group observed by Howard et al. We will return to this in \S~\ref{Hotties}.

Figures~\ref{Snap0}--\ref{Snap2.5} show the characteristic evolution of three different disks. All have the same
total mass ($50 M_{\oplus}$) within 1~AU, but distributed according to surface densities with different slopes. The disk with the flattest profile
($\alpha = 0$) is shown in Figure~\ref{Snap0}. In this case, the masses of the initial oligarchs increase rapidly with radius, scaling with the
area of individual annuli. The consequence of this is that the most massive planets at the end are
located from 0.5--1~AU. This is somewhat too large a radius to match most of the observations, which, at the time of writing, are most complete
for orbital periods of 100~days or less. Figure~\ref{Snap1.5} shows the evolution for
an intermediate slope ($\alpha=1.5$). Here we find that the most massive final planets are found between 0.2 and 0.5~AU. Although
the original oligarchs still increase in mass with semi-major axis, there is sufficient mass on smaller scales so that the faster
dynamical evolution results in the accumulation of mass into a few large, comparable-mass objects by 10~Myr. Figure~\ref{Snap2.5} shows the evolution of
a disk with a steep density profile ($\alpha=2.5$). In this case, the oligarch masses decrease with semi-major axis, and the final
result retains many of the characteristics of the original mass profile, with the largest planets having the shortest orbital periods. The three histories shown here demonstrate the characteristic
behaviour of the models -- steep density profiles result in a distribution of planet masses that decreases with increasing orbital
period, while flatter profiles move the maximum planet mass to larger scales. 
 The stochastic nature of N-body systems means that there is some variation in the total number and division
of mass from one simulation to the next, but the general distribution of mass with semi-major axis is reasonably constant. 

For a fixed density profile, the final planet masses will also, obviously, scale with the total mass budget. Figure~\ref{Param2_4}
shows the mass and orbital period of the largest final planet for several simulations with $\alpha=1.5$ (the choice which comes closest to
matching the observations). The open circles indicate a 
total disk mass of 100$M_{\oplus}$, the open triangles indicate $50 M_{\oplus}$ and the crosses indicate a total mass of $25 M_{\oplus}$.
The filled circles are the observed periods and masses of the largest bodies in systems detected by Howard et al. (2010) (Jupiter mass planets
will lie off the top of this plot). We see that the known super-earth candidates require total disk masses in the range $\sim 50-100 M_{\oplus}$.

As noted in \S~\ref{Inventory}, even the optimistic estimates of the static nebula solid inventory are not sufficient to
meet this requirement of $\sim 50$--$100 M_{\oplus}$ needed to match the observed planets.
Such massive planetesimal disks are more plausible if we allow for the possibility of migration of solids from several AU to small scales.

%The fact that the steeper profiles produce planets that populate the observed gap mass gap $20M_\oplus<m<100M_\oplus$ at $P\sim10$~days is a 
%demonstration of the fact that the planet accumulation process itself does not impose a natural scale at the observed edge. 
%On sub-AU scales, the planets are
%deep enough in the potential well of the star that the accumulation of disk material into planets is very efficient. In the outer solar
%system, the giant planets can eject significant amounts of material, some of which become the comet population. For short period
%planets, this only occurs when the masses approach Jupiter scales, well above the 10-50$M_{\oplus}$ scales under consideration
%here. The observed gap is therefore a consequence of the processes that regulate the amount of material available for planet formation.

\subsubsection{Quantitative Comparisons}
\label{Quant1}

Another generic consequence of in situ accumulation of rocky planets is that the resulting systems are
almost always multiple, and the degree of multiplicity and clustering can be characterised by a variety
of statistics (see Chambers 2001 for a summary, and appendix~\ref{Stats}) that were initially developed to analyse the formation
of the terrestrial planets of our own solar system. This proves a useful test of the in situ accumulation
hypothesis, as we would then expect the observed systems to be primarily multiple, and to be consistent
with the statistical distribution of the simulated systems\footnote{Multiplicity itself is not unique to
the in situ scenario, as migration models also generically produce multiple systems (e.g. McNeil \& Nelson 2010)
However, different scenarios should produce differences in the statistics of the resulting systems.}.
 A complication
for such a comparison is that the observations are likely to be still severely incomplete (Howard et al. 2010) and there may
be many additional planets lurking in the observed systems, but we can hope for this to improve with time.

However, there are some systems for which the noise levels are low enough to demonstrate the existence of multiple
systems already. The Howard et al sample contains several multiple systems, although some contain planets
beyond 1~AU, and so are not directly comparable to our calculations. However, there are two systems, 61 Vir (originally
announced by Vogt et al. 2009) and
HD69830 (originally announced by Lovis et al. 2006) which contain multiple planets within the period and mass range under discussion.
\footnote{There is an additional multi-planet system in this sample, which prompts some caution in advancing the above model
too broadly -- namely that of 55 Cancri. We defer discussion of this system until \S~\ref{Hotties}.}

Figures~\ref{Param2} and \ref{Param3} show the comparison of these two systems with the corresponding simulation results. In order to
ensure a proper comparison, we do not count planets in the simulations whose radial velocity amplitude would be $< 1 m/s$.
Figure~\ref{Param2} plots two statistical measures, $S_s$ and $S_c$, of how the mass in the system is distributed with semi-major axis. The observed systems HD69830
and 61 Vir are
shown as well , where the range indicates an uncertainty in the global system mass because of the unknown inclination
angle of the system to the line of sight, assuming coplanarity to within 5 degrees.
 Again, we see that
the $\alpha=3/2$ model produces systems with similar values of both mass concentration and distribution in eccentricity and semi-major axis, whereas the shallower and steeper slopes cannot do match all observed statistics simultaneously. To some
extent this is not surprising, given the qualitative features of Figures~\ref{Snap0}--\ref{Snap2.5}, although it is a more quantitative statement.
Figure~\ref{Param3} compares the measures of angular momentum deficit, AMD, (a measure of eccentricity and inclination spread)
and mass-weighted mean semi-major axis. The spread on the observed systems account for randomly varying the observed
eccentricities according to the observed error bars, and allowing the unknown mutual inclinations to fluctuate randomly by 5 degrees. Again,
we see that the $\alpha=3/2$ models bracket the observations.
 It is striking that both the 61 Virginis and HD69830 parameters overlap with the simulated values, suggesting that it
is possible to realise these observed systems within this model. The spread in the model statistics derives primarily from
two sources. The variation in the mass concentration statistics (Figure~\ref{Param2}) reflects the degree to which the mass
is spread evenly amongst the surviving planets, or concentrated in one or two more massive bodies. The spread in the
AMD (see appendix~\ref{Stats}) seen in Figure~\ref{Param3} is caused primarily by one of the surviving planets having
significant ($> 10^{\circ}$) inclination with respect to the others. If such conditions apply to the observed systems, then
their parameter range will be expanded from that shown here, in which we limited the amount of assumed mutual inclinations.
The outlier systems in the simulation results usually result from an incomplete accounting due to one or more of the final
bodies being too small to meet our radial velocity criterion.

The result of these accumulation models suggests that both the 61 Vir and HD69830 systems are consistent with in situ
accumulation of rocky planets from disks of sufficient mass. It also suggests that many of the other systems in the Howard et al
sample should harbour planets not far below the current detection limits, as the ratio of the two most massive planets
in the simulated systems ranges from 0.4--0.98, with a median value of 0.78. 

\section{Gas Accretion}
\label{Accretion}

The ability of the in situ accumulation model to match the properties of the observed low mass planet population is encouraging,
but the masses of the rocky bodies that result suggest that a reconsideration of the issue of gaseous accretion is warranted.
In the traditional model of gaseous giant planet formation, the first stage is the collisional assembly of a rocky core,
which eventually gravitationally captures gas from the surrounding nebula (e.g. Pollack et al. 1996) when it reaches as mass
$\sim 10 M_{\oplus}$. The final planet masses in \S~\ref{Accum} exceed this threshold, albeit in a much different region of
the disk than the original cores of Jupiter and Saturn. Therefore, the assumption that these planets are wholly rocky is likely
to be in error.

The question of gas giant formation on small scales was widely reconsidered upon the discovery of 51~Pegasi and other `hot Jupiters'.
However, a consensus quickly emerged that these planets most likely formed on larger scales and migrated to their currently observed
positions.
Part of this consensus was the finding, 
 from 
detailed calculations of the core accretion process, that in situ formation was unlikely
(Bodenheimer, Hubickyj \& Lissauer 2000; Ikoma, Emori \& Nakazawa 2001, Rafikov 2006).
However, these findings rest
largely
on the assumed implausibility of assembling a massive enough rocky core on small scales, given traditional estimates
of the rocky material reservoir. If the solid mass available  is substantially enhanced by inward radial transport of material, then
such massive cores are no longer an outlandish idea.

Furthermore, on these scales, 
 the assembly timescales of the planets in \S~\ref{Accum} is short, often $\sim 10^5$ years,
and  the system configurations are largely in place by $10^7$~years. Thus, it is quite conceivable that
cores of several earth masses are in place before the gas is removed from the system.
As a demonstration of this, 
Figure~\ref{Acc} shows the temporal history of the accumulation in a representative simulation with
total disk mass in solids of 50 $M_{\oplus}$ interior to 1~AU. 
 The final configuration has 4 planets in the `super-earth' mass range
interior to 1 AU. The assembly of all four planets is largely completed within 1~Myr and so it is quite
likely that the planets achieve a significant mass while there is still gas present in the system. 

On scales $\sim 0.1$~AU, detailed calculations suggest that the energy transport in a hydrostatic atmosphere
above a rocky core is regulated by convection (Ikoma et al. 2001; Rafikov 2006). 
However, these authors also note that traditional estimates of the critical core mass on these
scales become uncertain because the Hill sphere radii of the resulting cores begin to approach the
size of the disk scale height -- i.e. a  core could potentially open a gap in the disk before the
threshold mass is reached. For a disk temperature profile given by 
\begin{equation}
T = 280 K \left( \frac{a}{1 AU} \right)^{-1/2} \label{Tprof}
\end{equation}
 this is
$h \sim 0.05~{\rm AU} \, (a/1 AU)^{5/4} $. The criterion for the planet to open a gap in the gas disk is to
set the 
 Hill diameter  of a planet of mass $M_{core}$  equal to the gas disk scale
height $h$, which yields 
\begin{equation}
M_{core} \sim 1 M_{\oplus} \left( \frac{P_{orb}}{1 day} \right)^{1/2}. \label{GapMaker}
\end{equation}
This threshold mass is less than many of the simulated planets in \S~\ref{Accum} and most of the observed planets in the Howard et al. sample.
Thus, if the assembly of these planets takes place in the presence of gas, accretion is likely to occur. However, the final outcome will
not be a traditional gas giant planet. The normal core accretion sequence of events proceeds through a sequence of hydrostatically supported
atmospheres until runaway gas accretion occurs (see Rafikov 2006 for a review). However, for planets on these small scales, the hydrostatic 
atmosphere becomes tidally limited before instability sets in, and the opening of the gap limits the amount of gas accretion. Therefore, 
 tidally limited accretion from the disk will produce 
 a different, gas-starved, branch of the core accretion family than is found on larger scales.

For a gas disk of surface density $\Sigma \propto a^{-\beta}$, the gas mass within an annulus of half width equal to the Hill sphere of a core of mass $M_{core}$ is 
\begin{equation} \label{eqn: gap mass}
M_{gap} \sim 72 M_{\oplus} (2-\beta) \left( \frac{a}{1 AU} \right)^{9/4-\beta} \left( \frac{M_{gas}}{2.3 M_{J}} \right)
\end{equation}
where $M_{gas}$ is the gas disk mass inside 1 AU, and $a$ is the orbital semimajor axis of the core. The value $M_{gas}=2.3 M_{J}$ corresponds to the gas mass in the
standard Hayashi minimum mass solar nebula. The relation between this mass and the solid mass will depend on whether
or not solids migrate radially. If the gas is eventually accreted by the planet, then the 
 final mass of the planet will  result from the combination of the two estimates. In the case where both the
gas and solids follow a profile with $\alpha=\beta=3/2$, this yields a constant core mass fraction for the final planet,
\begin{equation}
\frac{M_{core}}{M_{tot}} = \left( 1 + 1.88 \frac{M_{gas}}{2.3 M_J} \right)^{-1} \label{GasScale}
\end{equation}
suggesting core mass fractions $\sim 0.15-0.35$, for ranges of disk mass from 1--3 the MMSN value. 
 If the gas disk had a different profile, the core mass fraction will 
vary with orbital period. For a gas disk profile $\beta=1$, the mass-period relation would be 
\begin{equation}
M = 1 M_{\oplus} \left( \frac{P_{orb}}{1 day} \right)^{1/2} + 0.53 M_{\oplus} \left( \frac{M_{gas}}{2.3 M_{J}} \right)
\left( \frac{P_{orb}}{1 day} \right)^{5/6}.  \label{GapLimit}
\end{equation}
It is striking that this estimate seems to naturally produce planets in the right range of masses and orbital periods. 
For example, at the fiducial period of 50 days, we derive a gap opening core mass of 7 $M_{\oplus}$ which results in a total mass of 21--49$M_{\oplus}$,
for gas disks within a factor of three of the MMSN. This naturally covers the range of the observed masses, with most
of the currently observed planets falling in this intermediate class with comparable amounts of rock and gas (see Figure \ref{PM}).

There is potentially also an intermediate regime, where the planet core is large enough to maintain a gravitationally
bound atmosphere but not big enough to open a gap in the gas disk. As the disk gas dissipates, the planet may 
retain enough of this gas to have a significant observational effect. This is the case treated by Rafikov (2006),
whose estimate of the core instability threshold is really a criterion for the gravitionally bound
gaseous atmosphere to be some fraction of the core mass. If we adopt an atmosphere fraction of 10\%, Rafikov's
estimate (in the case of the inner disk -- his equation~60), amounts to 
\begin{equation}
M \sim 1.5 M_{\oplus} \left( \frac{P_{orb}}{1 day} \right)^{5/12}.
\end{equation}
This turns out to be very similar to equation~(\ref{GapMaker}), so that there are probably few planets in the
intermediate regime. If a planet accretes a non-negligible mass fraction of gas, the accretion is likely limited
by the opening of a gap.

Thus, we envision two different outcomes for in situ rocky accretion, depending on the amount of mass available. Traditional
terrestrial class planets, such as those in the solar system, emerge from disks below some mass threshold so that the time to 
assemble bodies of several earth masses is longer than the gas dispersal timescale ($\sim 10^7$~years).
For larger mass disks,  cores form that are massive enough to bind gaseous atmospheres of significant mass before
the dispersal of the disk. This process is also self-limiting, as they will accrete enough gas to actually open a gap in
the disk, which limits their masses to the observed values. It does not depend sensitively on the evolution of the background
gas disk -- a difference from the traditional `failed core' model,
 often invoked to explain
these objects (e.g. Mordasini et al. 2009; Ida \& Lin 2010).
%We suggest that it is this population that is currently being detected in the
%latest generation of radial velocity surveys. 

\subsection{Assembly with Gas Accretion}

To determine the threshold of disk mass that divides the gas-poor and gas-rich systems, we have simulated the assembly of disks of various masses for 
an $\alpha=3/2$ profile, and find that the threshold disk mass is $\sim 25 M_{\oplus}$ of solids inside 1~AU. For disk masses above this 
value, one or more planets can assemble, within 1 Myr, 
to masses that satisfy equation~(\ref{GapMaker}), and therefore should accrete gas. For lower mass disks, planets either never reach the threshold value or
do so when the gas has dispersed, leaving behind rocky cores.  

Figure~\ref{PM} shows the overall picture that emerges, along with the observed planets in the Howard et al. sample. The
solid and open points indicate planets and planet candidates. The dotted line indicates the threshold mass for gas accretion (eqn. \ref{GapMaker}). 
Planets above this dotted line will frequently harbour non-negligible gaseous envelopes. The exact size of the envelope depends on the
mass in the gaseous disk. The long-dashed line indicates the final mass if the gas accreted is given by equation~(\ref{GapLimit}), and
assuming a gas disk that contains $9.5 M_J$ inside 1~AU (equivalent to four times the MMSN value). The region between these two
lines is approximately the domain of the Neptune-like planets born in situ. The short-dashed line running diagonally from upper left to lower 
right divides the parameter space into regions where accretion of solid bodies is efficient (to the left of the line) and not efficient (to the right).
 The in situ solid body accumulation process is largely conservative of mass on scales smaller than this, whereas planets that lie above the 
line are more likely to eject smaller bodies than to accrete them. These planets will not grow significantly, but rather will migrate inwards 
until they reach the accretion/ejection boundary given by the short-dashed line in figure~\ref{PM} (Murray et al. 1998). Further accretion will 
drive planets inwards approximately along the boundary (i.e. they
will grow and migrate inwards). The fact that there are few objects located near the intersection of these lines suggests that
concurrent migration and accretion in this fashion is not a significant contributor to the demographics of the observed populations.

%\subsection{Effect on the statistics}
%\label{Quant2}

%In \S~\ref{Quant1} we showed that the assembly of rocky planets in situ could match the observations of multi-planet
%systems in a variety of quantitative measures. However, if the rocky cores increase their masses by accreting gas, both
%the mass accounting and the dynamical rearrangement of the multi-planet system may be affected. Therefore, we now repeat
%the simulations with a simple model of the gas accretion process included.

The accretion of gas will change the assembly history, because one or more bodies will increase in mass and therefore the
gravitational perturbations on other bodies in the system  will be stronger. To account for the effects of this, we
 repeated the in situ assembly simulations using the $\alpha=3/2$ profile, but we stop the simulation at 0.1, 0.3 and 1~Myr to compare the masses
of all bodies to the threshold value of equation~(\ref{GapMaker}). If any body has crossed the threshold in the interim, we
instantaneously increase the mass according to a model for gas accretion, and then continue the simulation. This
model is not fully self-consistent as it does not treat the gravitational influence of the gas disk before it
is accreted, nor does it account for any difference in the specific angular momentum of the accreted gas, but it does
address the issue of how the rapid mass growth of some bodies affects the gravitational
interactions that regulate the  assembly of the N-body system into a
final system of planets. After the last gas accretion episode the systems are run to ages of 10~Myr.

We examined two models for gas accretion. In the first, we started with a rocky disk of mass $50 M_{\oplus}$, $\alpha=3/2$ and assumed gas
was accreted according to equation~(\ref{GapLimit}), i.e., from a gas disk that also obeys $\beta=3/2$, and with a mass three times the MMSN value. This gas 
disk mass is chosen to produce planets that trace the upper edge of the observed lower mass planet population. In the second model, 
we start with a lower mass rocky disk ($40 M_{\oplus}$, $\alpha=3/2$) and assume gas is accreted according to equation~(\ref{GasScale}), i.e., from
a gas profile $\beta=1$, 
with a MMSN normalisation. This is chosen to capture the median parameters of the observed
low mass population. Note that the gas and solid disks have different power law slopes, which is possible if the solid inventory was assembled
by radial inward migration, and not in situ deposition.

Figure~\ref{Param4} shows the statistical comparison of these simulations with the observed parameters of 61~Vir and HD69830. Both sets
of simulations show an inward shift in terms of the mass-weighted semi-major axis. This is a result of the tendency of planets with 
shorter orbital periods to assemble more rapidly and therefore they are more likely accrete gas. The end result is that including gas 
accretion in the model tends to steepen the mass profile. 

We have therefore repeated the experiments using a slightly shallower rocky profile of $\alpha=1$, using two different normalisations
(total masses of 50 $M_{\oplus}$ and 80 $M_{\oplus}$). In both cases, the gas accretion is given by equation~(\ref{GasScale}). 
The results are shown in Figure~\ref{Param7}. With the use of a steeper
profile, 
  both 61 Virginis and HD69830 remain within the parameter space of the
simulations, although the simulations are now biased to larger semi-major axis (because there is a larger fraction of mass 
at large radii and so it is easier to assemble the critical core mass). The overall conclusion to draw is that simulations
can produce a wide range of systems, depending on the original slope and the amount of gas accreted, bracketing those observed.

The dynamical effect of the gas accretion is to accelerate the orbital instabilities and dynamical interactions that characterise the
final stages of terrestrial planet assembly. Figure~\ref{TA22} shows an example of one system (drawn from the $\alpha=3/2$ case) in which three different planets
cross the threshold given by equation~(\ref{GapMaker}), two at 0.1~Myr, and a third at 0.3~Myr. Close examination of the
various curves shows that a rapid increase in mass leads to orbital instability in neighbouring, less massive bodies. The
ultimate result is frequent collisions, and occasionally accretion onto the star or ejection from the system. This example 
also shows a common occurrence in which two of the planets which accreted gas eventually merge. The end result of this
simulation is that we are left with three planets, two of which possess substantial gas fractions, although somewhat less
than indicated by equation~(\ref{GasScale}) because of the accretion of substantial amounts of rocky material during the
final clearing stages. This plot
also shows why the mass-weighted semi-major axis $<a>_M$ moves inwards for these simulations, relative to the purely rocky
ones -- the innermost planets grow the fastest and accrete gas earlier. This shifts the mass distribution to shorter
radii. Nevertheless, accretion occurs for planets at larger radii in some simulations as well, and we also find collisions
between planets on larger scales too. This offers an interesting avenue to get the upper envelope of the masses in Figure~\ref{PM}
even if the gas disk is not quite as massive as assumed in equation~(\ref{GapLimit}).

\section{Jovian Mass Planets}
\label{Hotties}

Our model thus far has addressed the nature of the low mass population at short periods. The Howard et al. (2010) 
sample also contains a handful of genuine Jovian mass planets with orbital periods $<$50~days. These must
arise in some other fashion. The original model for such planets was migration through a gas disk, but 
recent observations suggest the situation may be different. For planets with orbital periods of a few days, tidal
capture from eccentric orbits might be an alternative model (Rasio \& Ford 1996; Matsumura, Peale \& Rasio 2010). However, this seems less likely to produce 
a planet with orbital periods $\sim 10$~days, because tidal forces weaken rapidly with distance.

The planetary system around the star 55~Cancri may offer insights because it appears to contain members
of both the low mass and high mass population. We have investigated several hypotheses to see which are
consistent with the existence of the 55~Cancri system.

\begin{itemize}
 \item  Can the disk of rocky material we invoke play the role of a dissipative medium in capturing an eccentric
Jupiter? The model in which a high eccentricity planet is tidally captured was originally motivated by the onset
of dynamical instability, which can potentially occur late in a planetary system's lifetime (Rasio \& Ford 1996) after
the removal of the gas disk.
 We have therefore examined whether a rocky planet system
as described in previous sections can capture and circularise a Jovian mass planet injected into the system on
a highly eccentric orbit. The short answer is that this model is wildly unsuccessful, because the mass of the
injected planet is comparable to, or larger than, the entire mass budget of the disks we have considered. Thus,
the end result of this model is that the injected planet remains on an orbit with a slightly decreased semi-major
axis, and partially circularised eccentricity, 
 having ejected essentially all of the mass of the initial disk. To reduce the semi-major axis and eccentricity to the
level required by observation would require more mass, and would leave very little surviving mass to make the other
planets in the 55~Cancri system.\\
\item It is possible to generate rocky planets of Jovian mass, given a large enough mass reservoir.
In a model where the solid material is laid down from a marginally gravitationally stable $Q=1.5$ surface density profile (appendix~\ref{GI}),
the resulting innermost planet is quite massive, $\sim 100 M_{\oplus}$, with orbital periods $\sim 10$~days\footnote{At these masses, 
the conservative nature of terrestrial planet assembly starts to fail, see Figure~\ref{PM}}. This opens the possibility for the formation of
rocky planets of Saturn mass in short period orbits. At present there is little evidence for the existence of such objects in the sample discovered
in transit surveys, although that sample is heavily biased to shorter orbital periods, and is possibly dominated by the tidal capture of 
genuine gas giants from eccentric orbits. These simulations do not, however, faithfully reproduce the lower mass planets
found in the 55~Cancri system. Nevertheless, if some fraction of protoplanetary disks are significantly sculpted by gravitational
instability, this class of planets may eventually emerge in a sample like that of the Kepler mission.
\item A hybrid version of the model in \S~\ref{Accum} is to simply assume the giant planet was in place in the beginning through
some other, unspecified, process\footnote{One way to achieve this would be to apply the criterion of equation~(\ref{GapMaker}) in 
the case of the Q=1.5 gas disk, which
is much more massive. However, the rest of the 55 Cancri system is consistent with a less massive disk.}, and then to repeat our 
calculation of in situ accumulation with this additional component. The end result of such a simulation 
 actually produces systems qualitatively similar to that of 55~Cancri, as shown in Figure~\ref{55match}. We are able to produce
both an inner planet and a system of planets further out. In the event of no gas accretion, the resulting planets are too small compared
to the observed planets, but they do cross the threshold of equation~(\ref{GapMaker}) and should grow to appropriate mass. Figure~\ref{55match}
shows this for the analogue of 55~Cancri~c. The conclusion is that the presence of a jovian mass planet in the disk does not 
impede the in situ assembly of planets along the lines we have investigated, and that the resulting dynamical environment can produce
systems with similar organisation to 55~Cancri.
\end{itemize}

Jovian mass planets can also potentially influence the assembly on sub-AU scales even if they orbit farther out (such perturbations are
often advanced as potential causes for the low mass of the Asteroid belt, and possibly Mars -- e.g. Chambers 2001). There are several
systems in the Howard et al. sample which contain a gas giant at long orbital periods and a lower mass planet at shorter orbital periods.
To investigate the degree to which an external Jupiter could influence the assembly histories considered here, 
 we have repeated the $\alpha=3/2$, 50 $M_{\oplus}$ runs in the presence of a planet of mass 351~$M_{\oplus}$,
 orbital period 493~days, and $e=0.1$. This is chosen based on the planet HD114783b, as an example.
 We find that planet
assembly is largely unaffected interior to $\sim 0.65$~AU (orbital periods $\sim 180$~days). The bulk of our analysis applies to smaller scales and
shorter periods, so that the general character of the solutions is largely unaffected by the presence or absence of Jupiters on scales $>1$~AU.
This is consistent with the approximate separation between Jupiter and the asteroid belt in the solar system and the rule-of-thumb that a gas
giant does not significantly reduce planet assembly interior to it's 3:1 mean motion resonance.

\section{Evaporation}
\label{Evap}

It has been suggested by numerous authors that short period planets may experience significant
atmospheric mass loss driven by irradiation from the central star (e.g. Lammer et al. 2003; Yelle 2004; Lecavelier des Etangs et al. 2004; Tian 
et al. 2005; Murray-Clay, Chiang \& Murray 2009). As a result, we need to consider how much of a limit this will place on the accretion or
retention of gas by planets at short orbital periods.

Empirical estimates of the evaporation rate for exoplanets are based on the so-called `Energy Diagram',
in which the gravitational binding energy of the planet ($E_p$) is compared to the energy input ($L_{evap}$) from X-ray
and EUV radiation (Lecavelier des Etangs 2007, Ehrenreich \& Desert 2011), after accounting for various
uncertain efficiency factors. Using the estimates of Ehrenreich \& Desert for the life-time averaged
EUV flux for the host star, we can rewrite the formula for the characteristic evaporation time,
$T_{evap} = E_p/<L_{evap}>$, as a mass-period relation 
\begin{equation}
M_{evap} = 51 M_{\oplus} \left( \frac{T_{evap}}{5 Gyr} \right)^{1/2} \left( \frac{\eta}{1} \right)^{1/2}
\left( \frac{R_p}{0.1 R_J} \right)^{3/2} \left( \frac{P}{1 day} \right)^{-2/3} \label{EvapLim}
\end{equation}
where $\eta$ is the efficiency factor for converting incident radiation into mass loss (set equal to unity
here, so that this is a very generous estimate). The meaning of
this equation is that planets with masses above this threshold are too large to experience significant
dimunition through the evaporation process, under even very optimistic assumptions. 
 We see that this
relation has the wrong slope to explain the edge in Figure~\ref{PM}, and will likely affect only the
planets with periods $<10$~days. The slope may be changed if we postulate a particular relationship
between mass and radius, but not by much, as the mass-radius relationship for planets with gaseous
envelopes is relatively flat in this mass range. Therefore, evaporation is
 unlikely to be responsible for the shape of the mass-period relation in Figure~\ref{PM}. Nevertheless, it
 is likely to be important for the lowest mass, shortest period planets. In cases where the
mass given by equation~(\ref{EvapLim}) is greater than that given by equation~(\ref{GapLimit}), the planet
may very well accrete some gas, but it will likely lose it again over the course of time. Thus, we may
expect to see a transition from significant gaseous envelopes to thin, remnant envelopes below some period.
Estimating the period is somewhat uncertain because of the rather unconstrained nature of $\eta$, but it
likely lies in the range of 3.7~days ($\eta=0.1$) to 8.5~days ($\eta=1$).

\section{Discussion}

Motivated by recent observational challenges to the prevailing paradigm of exoplanet migration, we have
re-examined the role that in situ accumulation might play in the assembly of planets on small scales,
with masses ranging up to several tens of earth masses. We 
 have demonstrated that the observed properties of planets in 
the ``Hot Neptune"/''Super-Earth" mass range, orbiting solar mass stars with periods of days to several months,
can be reproduced with a simple model of in-situ accumulation of rocky planets from traditional oligarchic
conditions, without any additional radial migration. However, in
 order for this to be the case, the original amount of rocky material required is $\sim 30-100 M_{\oplus}$
within 1~AU. This is somewhat in excess of that expected from traditional estimates of the mass profiles
of protoplanetary disks, but could be achieved if there was significant radial migration of solid material from
the outer parts of disks to the scales of interest, before the migrating bodies achieved the mass of planetary
embryos. Thus, our calculations argue that migration of solid material did indeed occur, but that it served to
rearrange the mass budget of the nebula before the final stages of planet assembly.
Such a position is not without physical foundation, as the timescale for small bodies to migrate over AU scales
 through the
action of aerodynamic drag forces can be as small as $\sim 100$~years (Weidenschilling 1977b; Nakagawa et al. 1986; Youdin \& Shu 2002;
Bai \& Stone 2010). 

A further consequence of such a mass concentration is that the collisional assembly of planets is fast enough that bodies
of several earth masses can form within 1~Myr of the original material being emplaced. If this growth occurs
in the vicinity of a gas disk, accretion and gravitational retention of gas from the nebula is possible. The
accretion of this gas is limited, however, by the fact that it is easier to open a gap in the gas disk on
smaller scales, and the accretion may be quenched before the accretion flow enters the traditional runaway
regime in the core instability scenario (e.g. Rafikov 2006). This process provides a natural mechanism for the
production of planets with comparable amounts of gas and rock, intermediate between the traditional classifications
of rocky terrestrial planets, and gas-dominated giants.

Figure~\ref{MX} shows the heavy element fraction of planets of various mass formed in our simulations, compared
to published estimates for observed 'Hot Neptunes' around solar mass stars. It should be noted that many
of the published systems do not strictly fall within the bounds of our model (e.g. HD149026b) in terms of
the parameter space of Figure~\ref{PM}. Furthermore, our crude estimate of a tidally-limited gas accretion 
ignores a lot of potentially interesting physics that determines the final mass of a planet embedded in a gas
disk. In particular, planets that open gaps in disks can continue to accrete gas if the disk viscosity is high
enough to allow flow into the gap (e.g. Artymowicz \& Lubow 1996; Bryden et al. 1999; Kley 1999). Based on the
formulae of Lin \& Papaloizou (1979) or Taniguchi \& Ikoma (2007), we estimate that disks with $\alpha \sim 10^{-3}$
(Shakura \& Sunyaev 1973) may result in further gas accretion across the gaps we have estimated here.
 Nevertheless, it is interesting that the simple considerations outlined above provide
a natural framework for producing planets of this observed class. It is particularly encouraging that the objects
 on this plot most likely to fall within our model are the Kepler-9 planets, which match the results of our simulations quite well.

The quantitative agreement between observations and model is in striking contrast to the failures of
traditional migration-based models (Alibert et al. 2006, Terquem \& Papaloizou 2007; Ida \& Lin 2010) to match the same observations
(McNeil \& Nelson 2010). The Neptune mass analogs in these calculations formed further out and represent `failed cores' of giant
planets, whose inward migration due to disk torques quenches the gas accretion that would produce a traditional gas giant.
Multiplicity is often important in the survival of such objects, as strong mutual gravitational interaction can serve to slow
the migration and avoid spiralling into the central star. However, the effects of such strong interaction imprints itself on
the final system, even after the loss of gas and further dynamical evolution (Ida \& Lin 2010; McNeil \& Nelson 2010).

 Recent observations of hot Jupiters have cast doubt on the migration of gas giants
as well, suggesting that perhaps the very idea of planetary migration needs to be extensively reviewed (although
we stress that the inventory of solids invoked in our model most likely still requires significant radial
migration of solid material). We should be clear, however, that the model presented here is not complete, as
we have not addressed the details of how these systems are related to those with hot Jupiters or Jupiters on
more traditional scales. Our aim is to better understand the individual physical processes at work before attempting
a complete synthesis of all known planetary systems. Furthermore, it is now evident that the lower planet mass systems
are the most generic kind, and that stars which host hot Jupiters are a small minority. It may thus be useful to
consider our models independently of those developed for hot Jupiters, which could represent only one very special case amongst
the pathways to planetary system formation.

Our original discussion was purposely limited to the sample of Howard et al. (2010) because of the homogeneity of selection.
Mayor et al. (2011) recently released a catalog of detections from a homogeneously selected sample of their own. In addition
to 61~Vir and HD~69830, their sample contains seven more stars with at least three sub-Jupiter mass companions.
 Figure~\ref{Param6} shows a comparison of these systems with 
 the same models as discussed before. The model range encompasses nearly all of these, with HD~10180 (Lovis et al. 2011),
HD~31527 and HD~136352 located closest to the bulk of the simulation points. Several systems, such as HD~40307 and HD~39194, could be
brought closer to the main locus of points if additional planets are discovered in these systems with longer orbital period (which
seems likely in the case of HD~40307, since Mayor et al. 2009a also report the detection of a linear trend).
Conversely, HD~134606 has an anomalously large  $\langle a\rangle_M$, which could be made consistent if another planet
is found in the system with orbital period $\sim 150$~days. 

%The three most extreme outliers are Kepler-11, Gliese~876, and Gliese~581. The latter two are systems that
%orbit lower mass M~stars, so that they are not strictly comparable to the simulations above. Furthermore, the
%Gliese~876 system (Marcy et al. 1998; 2001; Rivera et al. 2005) contains two Jovian-mass planets, suggesting that other processes of formation may apply.
%In the case of the Gliese~581 system (Bonfils et al. 2005; Udry et al. 2007; Mayor et al. 2009b; Vogt et al. 2010) , the $S_s$ value 
%is anomalously low, but could be brought more in line
%if the outermost planet has a somewhat overestimated mass. Given the controversy over the recently published
%parameters in this system, this is not an outrageous possibility (the putative habitable planet is not massive
%enough to significantly affect these parameters). The Kepler-11 system parameters (Lissauer et al. 2011b) can also be brought more into
%line as the (poorly known) outer planet mass estimate is revised downward. 

As a check on the Howard et al. sample, 
Figure~\ref{pmall} shows the HARPS (Mayor et al. 2011) and Kepler-based exoplanet samples in the orbital period-mass plane, along with the
proposed physical classification based on our model. At the lower right, at masses below those given by equation~(\ref{GapMaker}), 
planets are not large enough to accrete significant gas and thus are likely to be truly rocky planets. This is a regime that is poorly
explored as yet, although current radial velocity limits are slowly inching lower. Above this limit, and at orbital periods
$> 10$~days, we expect to find cores which have accreted substantial, but not overwhelming, gaseous envelopes. Within our
model, the bulk of the planets in this period and mass range are expected to be more like Neptunes than scaled-up rocky
planets (a.k.a. Super-Earths). This is consistent with the newly emerging constraints from Kepler, which suggests planetary
radii for this class of object that require substantial gaseous envelopes (Holman et al. 2010;  Lissauer et al. 2011b).
Further encouragment comes from the analysis of Howard et al. (2011), who estimate that planets with masses $> 4.5 M_{\oplus}$ and
periods $< 50$~days have substantial gas fractions on average, while those at lower masses are less gas-rich. Evaluating
our equation~(\ref{GapMaker}) at a mean period for the Howard sample $\sim 25$~days, yields a mass of 5$M_{\oplus}$. As
the observations and analysis improve, we expect that this division will be found to rise to higher masses at longer orbital periods. Although the
criterion is based on the ability of a core to gravitationally retain gas (e.g. Ikoma et al. 2001; Rafikov 2006), if the
ordering of mass with semi-major axis is preserved to the present day, it would represent strong evidence that there was
little orbital evolution post assembly.
 For orbital periods $< 10$~days, the division between the two classes is less uncertain because of the possibility that
a gaseous envelope could be accreted but then removed over time. The fact that the observations tend to group near the line
in Figure~\ref{pmall} is a tantalising suggestion that evaporation could be very efficient. If true, it would suggest that
most of the planets with $P< 10$~days, and $1 M_{\oplus} < M < 10 M_{\oplus}$ are Neptune-like planets that have experienced
significant erosion over the course of their lifetime.

The essential point of these calculations is that many of the quantitative features of the observations can be modelled within
a relatively simple in situ accumulation formalism. As such, several aspects of the more general astrophysical context need to
be addressed, but the agreement with the observations argues that the basic model is correct.  The mechanism(s) by which heavy
elements are concentrated in the inner parts of the nebula remain to be properly quantified. Although the mobility of small
bodies in protoplanetary nebulae is well established, the precise nature of the material rearrangement will depend on the properties
of the background nebula. Similarly, the effects of a residual gas disk on the damping of planetary eccentricities and inclinations 
is something to be considered. However, Moeckel \& Armitage (2012) have shown that the statistical outcome of planetary scattering
remains largely unaffected by the presence of gas. They do identify a subpopulation of low eccentricity orbits in their population, but
presently the observations are still in excellent agreement with the outcomes of pure N-body calculations (e.g. Figure~\ref{pmall}).
The model also predicts a period-dependent threshold mass that separates true rocky planets from those with comparable amounts of rock
and gas. This will be quantitatively tested as more low mass planets are found, but already finds some statistical support in the
planet candidates uncovered by Kepler (Howard et al. 2011; Wolfgang \& Laughlin 2011; Youdin 2011).

\acknowledgements  The authors thank Phil Armitage, Ravit Helled, Kristen Menou, Hilke Schlichting and Chris Thompson for discussions and
comments. B.H. is supported by the NASA ATP program. N.M. is supported in part by the Canada Research Chair
program and by NSERC of Canada.

\newpage

\appendix

\section{Hybrid Oligarch Model}
\label{Democratic}

For the kinds of disk masses considered in this paper, the oligarchic formalism of Kokubo \& Ida (1998) can
produce bodies with masses $10 M_{\oplus}$ or more. With masses this large, one might wonder whether the 
long-range gravitational perturbations between neighbouring annuli become important sooner, and perhaps the
assumption of initially independent annuli no longer holds. To examine this, we have repeated our calculations
but have restricted the maximum size of an oligarch to $1 M_{\oplus}$. The bodies are still spaced by annuli
with width ten times the Hill radius, and so the overall surface density distribution is preserved. In Table~\ref{Democheck}
we show the statistics for 5 realisations of this model, for a disk with power law $\alpha=1.5$ and total mass $50 M_{\oplus}$.
These can be compared with the results of the runs Run$\_{50}\_{1.5}\_1$--12 in Table~\ref{SimParam}. 

We see that there is little difference in the final statistical measures. The only noticeable difference is that 
the runs for this section lose a few percent of material in scattered and ejected bodies, which leads to a slightly
higher probability that one of the surviving bodies will have a radial velocity signature $< 1 m/s$. As a result we
quote two sets of statistics for some of the runs in Table~\ref{Democheck}. The first line represents the direct comparison
to other simulations. In those cases where we show a second set of values for a simulation, it means that one of the 
surviving planets would not have been detected with a 1 m/s radial velocity threshold. The second line shows the statistics
calculated if we include those planets that would likely have been missed.

\section{Statistical Measures}
\label{Stats}

The equations collected here offer a brief description of each of the statistical measures used in the text. Further
information can be found in Chambers (2001) and references therein.
 The angular momentum deficit is defined as
\begin{equation}
AMD= \frac{\sum_j m_j \sqrt{a_j} \left[ 1 - \sqrt{(1-e_j^2) \cos i_j}\right] }{\sum_j m_j \sqrt{a_j}}
\end{equation}
where $m_j$, $a_j$, $e_j$ and $i_j$ are the mass, semi-major axis, eccentricity and inclination of the
j-th particle. It is a measure of how much the orbits in the system deviate from the case of
circular, coplanar orbits, which represent the state of maximal angular
momentum per unit energy.

The mass-weighted semi-major axis is
\begin{equation}
<a_M> = \frac{\sum_j a_j m_j}{\sum_j m_j},
\end{equation}
and is a measure of where the mass is eventually accumulated.

The mass concentration statistic is given as the maximum value of the function 
\begin{equation}
S_c = max \left( \frac{ \sum_j m_j}{\sum_j \left[ \log_{10} (a/a_j) \right]^2} \right)
\end{equation}
as the  semi-major axis $a$ is varied.

The orbital spacing statistic is
\begin{equation}
S_s = \frac{6}{N-1} \left( \frac{a_{max}-a_{min}}{a_{max}+a_{min}} \right) \left( \frac{3 M_*}{2 \bar{m}} \right)^{1/4}
\end{equation}
where $\bar{m}$ is the mean mass of the planets, and $a_{min}$($a_{max}$) is the semi-major axis of the shortest (longest) period of
the N planets in the system. This quantity is often considered a measure of the long-term dynamical
stability of the system, as it bears a qualitative similarity to the average spacing of the planets measured in terms of
their Hill radii, $\Delta$. Dynamical simulations can determine the threshold value of these parameters for long term
stability (e.g. Chambers, Wetherill \& Boss 1996), but require some choices with regards to the masses of the various
components. Thus, we have repeated this calculation for the specific case of interest here. We consider three-planet systems,
with the middle planet at 0.35~AU. The planet masses are given by equation~(\ref{GasScale}), and two other planets in the
system were placed at equal separation $\delta a$ on either side. The results suggest that stability is achieved for
$\delta a \geq 0.09$~AU, corresponding to $S_s>10$. Comparison to Figure~\ref{Param6} shows that nearly all the observed
systems satisfy this criterion. The two that do not are Gliese~876 (which corresponds to a pair of Jupiter mass planets orbiting
an M-dwarf in resonance) and Kepler-11 (which satisfies this criterion if the mass of the outermost planet, Kepler-11g, lies
somewhat below the currently quoted upper limit).

\section{Gravitationally Unstable Disks}
\label{GI}
In the early stages of star formation, the mass fraction of material in the disk may be substantial and the disk may exhibit a variety of large
scale gravitational instabilities. This can also lead to significant radial transport of gas and dust (e.g. Laughlin \& Rozyczka 1996; Gammie 2001) 
and may result ina somewhat different radial profile than those considered above. Thus, we will also consider a model in which the solid distribution is deposited
in situ, but from a gas disk which follows a disk profile dictated by setting the Toomre parameter to $Q=1.5$, which we take as the marginally stable
end state of a gravitationally unstable disk. The temperature profile is also somewhat different because the disk is heated by the dissipation inher
ent in
the mass transport. Based on the results of Rice \& Armitage (2009), we assume a constant temperature $T=1000$~K interior to 1 AU, and derive
\begin{equation}
\Sigma_Z = 404 {\rm g\,cm}^{-2} \left( \frac{a}{1 AU} \right)^{-3/2}  \left( \frac{Q}{1.5} \right)^{-1} \label{Qsig}
\end{equation}
where we have used a conversion factor of $10^{-3}$ between gas and solid material. This is a small fraction of the gas metallicity, but captures the
likely inefficiency of sedimentation at these temperatures.

The assembly of embryos, drawn from this distribution, into planets proceeds in the same way as in the other simulations, except for the
fact that there is considerably more mass available. The final planets produced have masses that range up to $100 M_{\oplus}$. This particular
scenario may never be realised in practice, but is worth calculating if only to demonstrate that the gap in the observed masses is the
result of the available reservoir of material, not of the physics of assembly itself. It is only when the masses begin to approach
the ejection/accretion division shown in Figure~\ref{PM} that the process of collisional accumulation becomes self-limiting.

\newpage

\begin{deluxetable}{lccccc|cccccc}
\tablecolumns{12}
\tablewidth{0pc}
\tablecaption{Simulation Inputs and Outputs. A simulation named Run$\_{X}\_{Y}\_{n}$ is the nth realisation of a
disk of mass X $M_{\oplus}$, with a surface density power law index $\alpha=Y$. Simulations with a `g' appended include
gas accretion according to the prescription described in the text. Simulations with an `N' appended include the gravitational
perturbations of a Jovian mass planet in a 493~day orbit, as described in the text. Statistics are calculated only for those
planets with semi-major axis $< 1.1$~AU and with a radial velocity amplitude larger that 1 $m/s$.
\label{SimParam}}
\tablehead{ 
 &
\multicolumn{4}{c}{Inputs}& & &\multicolumn{5}{c}{Outputs} \\
\colhead{Name} & \colhead{$M_Z$}   & \colhead{$\alpha$}  & \colhead{$M_{gas}$} & \colhead{$\beta$} & \colhead{$N_{tot}$} &
\colhead{N($<1.1 AU$)} &  \colhead{$M_{big} $} & \colhead{$S_s$} & \colhead{AMD} & \colhead{$S_{c}$} & \colhead{$<a>_M$} \\
  & \colhead{($M_{\oplus}$)} & & \colhead{($M_{\oplus}$)} & & & & \colhead{($M_{\oplus}$)} & & & & \colhead{(AU)} }
\startdata
Run$\_Q=1.5$ & 230 & 1.75 & $\cdots$ & $\cdots$ & 57 & 2 & 99.3 & 36.5 & 0.073 & 8.5 & 0.168 \\
Run$\_Q=1.5$ & 230 & 1.75 & $\cdots$ & $\cdots$ & 57 & 2 & 127.5 & 31.1 & 0.012 & 10.3 & 0.291 \\
Run$\_Q=1.5$ & 230 & 1.75 & $\cdots$ & $\cdots$ & 57 & 2 & 97.9 & 29.4 & 0.027 & 13.5 & 0.161 \\
Run$\_{25}\_{0}\_1$ & 25 & 0 & $\cdots$ & $\cdots$ & 79 & 1 & 16.3  & $\cdots$ & 0.034 & $\infty$ & 0.545 \\
Run$\_{25}\_{0}\_2$ & 25 & 0 & $\cdots$ & $\cdots$ & 79 & 1  & 12.5 & $\cdots$ & 0.1011 & $\infty$ & 0.485 \\
Run$\_{25}\_{0}\_3$ & 25 & 0 & $\cdots$ & $\cdots$ & 79  & 1  & 9.9 & $ \cdots$ & 0.0005 & $\infty$  & 0.906  \\
Run$\_{50}\_{0}\_1$ & 50 & 0 & $\cdots$ & $\cdots$ & 57 & 2 & 46.1  & 38.6 & 0.0136 & 22.8 & 0.740 \\
Run$\_{50}\_{0}\_2$ & 50 & 0 & $\cdots$ & $\cdots$ & 57 & 3 & 27.1 & 21.2 & 0.0084 & 17.3 & 0.738 \\
Run$\_{50}\_{0}\_3$ & 50 & 0 & $\cdots$ & $\cdots$ & 57 & 3 & 27.1 & 17.5 & 0.0031 & 45.0 & 0.705 \\
Run$\_{50}\_{0}\_4$ & 50 & 0 &  $\cdots$ & $\cdots$ & 57 & 3 & 42.7 & 26.3 & 0.0081 & 15.8 & 0.717 \\
Run$\_{50}\_{0}\_5$ & 50 & 0 & $\cdots$ & $\cdots$ & 57 & 3 & 32.1 & 15.6 & 0.0013 & 59.9 & 0.734 \\
Run$\_{50}\_{0}\_6$ & 50 & 0 & $\cdots$ & $\cdots$ & 57 & 3 & 28.0 & 11.4 & 0.0006 & 87.1 & 0.750 \\
Run$\_{50}\_{0}\_7$ & 50 & 0 &  $\cdots$ & $\cdots$ & 57 & 2 & 28.0 & 21.5 & 0.0014 & 43.3 & 0.767 \\
Run$\_{50}\_{0}\_8$ & 50 & 0 & $\cdots$ & $\cdots$ & 57 & 1 & 40.5  & $\cdots$ & 0.0194 & $\infty$ & 0.559 \\
Run$\_{50}\_{0}\_9$ & 50 & 0 & $\cdots$ & $\cdots$ & 57 & 3 & 28.0 & 21.9 & 0.0148 & 16.4 & 0.755 \\
Run$\_{100}\_{0}\_1$ & 100 & 0 & $\cdots$ & $\cdots$ & 41 & 2 & 61.3 & 23.4 & 0.0125 & 46.4 & 0.664 \\
Run$\_{100}\_{0}\_2$ & 100 & 0 & $\cdots$ & $\cdots$ & 41 & 3 &  51.3 & 19.3 & 0.0020 & 37.2 & 0.628 \\
Run$\_{100}\_{0}\_3$ & 100 & 0 & $\cdots$ & $\cdots$ & 41 & 4  & 50.3 & 18.5 & 0.0042 & 23.9 & 0.614 \\
Run$\_{25}\_{1.5}\_1$ & 25 & 1.5 & $\cdots$ & $\cdots$ & 34 & 5 & 6.3 & 24.1 & 0.0114 & 6.2 & 0.401 \\
Run$\_{25}\_{1.5}\_2$ & 25 & 1.5 & $\cdots$ & $\cdots$ & 34 & 3 & 10.2 & 36.6 & 0.0189 & 11.7 & 0.234 \\
Run$\_{25}\_{1.5}\_3$ & 25 & 1.5 & $\cdots$ & $\cdots$ & 34 & 4 & 9.3 & 29.4 & 0.0136 & 6.9 & 0.380 \\
Run$\_{50}\_{1.5}\_1$ & 50 & 1.5 & $\cdots$ & $\cdots$ & 25 & 3 & 17.3 & 32.1 & 0.0102 & 6.2 & 0.383 \\
Run$\_{50}\_{1.5}\_2$ & 50 & 1.5 & $\cdots$ & $\cdots$ & 25 & 2 & 16.5 & 48.5 & 0.0417 & 11.3 & 0.156 \\
Run$\_{50}\_{1.5}\_3$ & 50 & 1.5 & $\cdots$ & $\cdots$ & 25 & 2 & 23.3 & 48.7 & 0.0484 & 12.8 & 0.199 \\
Run$\_{50}\_{1.5}\_4$ & 50 & 1.5 & $\cdots$ & $\cdots$ & 25 & 3 & 22.6 & 33.7 & 0.0033 & 5.1 & 0.513 \\
Run$\_{50}\_{1.5}\_5$ & 50 & 1.5 & $\cdots$ & $\cdots$ & 25 & 4 & 19.9 & 24.6 & 0.0100 & 5.9 & 0.474 \\
Run$\_{50}\_{1.5}\_6$ & 50 & 1.5 & $\cdots$ & $\cdots$ & 25 & 3 & 17.8 & 33.3 & 0.0059 & 8.4 & 0.304 \\
Run$\_{50}\_{1.5}\_7$ & 50 & 1.5 & $\cdots$ & $\cdots$ & 25 & 4 & 11.7 & 30.4 & 0.0569 & 10.0 & 0.208 \\
Run$\_{50}\_{1.5}\_8$ & 50 & 1.5 & $\cdots$ & $\cdots$ & 25 & 3 & 26.4 & 31.9  & 0.0395 & 7.4 & 0.341 \\
Run$\_{50}\_{1.5}\_9$ & 50 & 1.5 & $\cdots$ & $\cdots$ & 25 & 4 & 15.9 & 34.0 & 0.0087 & 5.2 & 0.428 \\
Run$\_{50}\_{1.5}\_10$ & 50 & 1.5 & $\cdots$ & $\cdots$ & 25 & 4 & 12.8 & 24.6 & 0.0094 & 8.6 & 0.286 \\
Run$\_{50}\_{1.5}\_11$ & 50 & 1.5 & $\cdots$ & $\cdots$ & 25 & 3 & 23.3 & 33.9 & 0.0121 & 8.0 & 0.316 \\
Run$\_{50}\_{1.5}\_12$ & 50 & 1.5 & $\cdots$ & $\cdots$ & 25 & 4 & 14.5 & 24.2 & 0.0076 & 6.4 & 0.384 \\
Run$\_{100}\_{1.5}\_1$ & 100 & 1.5 & $\cdots$ & $\cdots$ & 18 & 5 & 28.3 & 16.8 & 0.0034 & 6.5 & 0.426 \\
Run$\_{100}\_{1.5}\_2$ & 100 & 1.5 & $\cdots$ & $\cdots$ & 18 & 2 & 53.3 & 54.5 & 0.0067 & 13.6 & 0.588 \\
Run$\_{100}\_{1.5}\_3$ & 100 & 1.5 & $\cdots$ & $\cdots$ & 18 & 4 & 37.3 & 20.5 & 0.0159 & 5.9 & 0.442 \\
Run$\_{50}\_{2.5}\_1$ & 50 & 2.5 & $\cdots$ & $\cdots$ & 20 & 3 & 24.0 & 26.5 & 0.0093 & 13.8 & 0.132 \\
Run$\_{50}\_{2.5}\_2$ & 50 & 2.5 & $\cdots$ & $\cdots$ & 20 & 3 & 19.4 & 30.9 & 0.0063 & 9.9 & 0.155 \\
Run$\_{50}\_{2.5}\_3$ & 50 & 2.5 & $\cdots$ & $\cdots$ & 20 & 3 & 22.6 & 28.0 & 0.0125 & 16.1 & 0.123 \\
Run$\_{50}\_{2.5}\_4$ & 50 & 2.5 & $\cdots$ & $\cdots$ & 20 & 4 & 19.8 & 19.4 & 0.0057 & 12.2 & 0.116 \\
Run$\_{50}\_{2.5}\_5$ & 50 & 2.5 & $\cdots$ & $\cdots$ & 20 & 4 & 19.8 & 18.7 & 0.0040 & 14.8 & 0.102 \\
Run$\_{50}\_{2.5}\_6$ & 50 & 2.5 & $\cdots$ & $\cdots$ & 20 & 5 & 16.6 & 17.3  &  0.0082 & 10.7 & 0.126  \\
Run$\_{50}\_{2.5}\_7$ & 50 & 2.5 & $\cdots$ & $\cdots$ & 20 & 4 & 14.5 & 20.7 & 0.0023 & 13.8 & 0.110 \\
Run$\_{50}\_{2.5}\_8$ & 50 & 2.5 & $\cdots$ & $\cdots$ & 20 & 4 & 14.5 & 21.2 & 0.0068 & 10.2 & 0.131 \\
Run$\_{50}\_{2.5}\_9$ & 50 & 2.5 & $\cdots$ & $\cdots$ & 20 & 4 & 13.4 & 21.3  &  0.0039 & 11.8 & 0.133  \\
Run$\_{100}\_{2.5}\_1$ & 100 & 2.5 & $\cdots$ & $\cdots$ & 14 & 4 & 40.5 & 21.0 & 0.0132 & 6.9 & 0.198 \\
Run$\_{100}\_{2.5}\_2$ & 100 & 2.5 & $\cdots$ & $\cdots$ & 14 & 3 & 60.3 & 25.2 & 0.0099 & 9.3 & 0.165 \\
Run$\_{100}\_{2.5}\_3$ & 100 & 2.5 & $\cdots$ & $\cdots$ & 14 & 4 & 49.7 & 18.6 & 0.0133 & 9.6 & 0.164 \\
\hline \\
Run$\_{50}\_1.5\_1g$ & 50 & 1.5 & 2300 & 1.5 & 25 & 4 & 62.0 & 19.6 & 0.023 & 11.1 & 0.311 \\
Run$\_{50}\_1.5\_2g$ & 50 & 1.5 & 2300 & 1.5 & 25 & 4 & 60.7 & 20.6 & 0.011 & 10.9 & 0.267 \\
Run$\_{50}\_1.5\_3g$ & 50 & 1.5 & 2300 & 1.5 & 25 & 4 & 101.4 & 17.6 & 0.0047 & 9.6 & 0.354 \\
Run$\_{50}\_1.5\_4g$ & 50 & 1.5 & 2300 & 1.5 & 25 & 3 & 40.8 & 27.9 & 0.0230 & 6.8 & 0.315 \\
Run$\_{50}\_1.5\_5g$ & 50 & 1.5 & 2300 & 1.5 & 25 & 3 & 67.3 & 23.5 & 0.0091 & 12.8 & 0.230 \\
Run$\_{50}\_1.5\_6g$ & 50 & 1.5 & 2300 & 1.5 & 25 & 3 & 52.0 & 27.6 & 0.0146 & 6.8 & 0.301 \\
Run$\_{50}\_1.5\_7g$ & 50 & 1.5 & 2300 & 1.5 & 25 & 3 & 42.6 & 27.9 & 0.0145 & 8.1 & 0.259 \\
Run$\_{50}\_1.5\_8g$ & 50 & 1.5 & 2300 & 1.5 & 25 & 2 & 100.0 & 19.9 & 0.0059 & 34.3 & 0.410 \\
Run$\_{40}\_1.5\_1g$ & 40 & 1.5 & 767 & 1.0 & 30 & 4 & 22.1 & 23.9  & 0.0025 & 9.3 & 0.217 \\
Run$\_{40}\_1.5\_2g$ & 40 & 1.5 & 767 & 1.0 & 30 & 3 & 15.7 & 32.3 & 0.011 & 12.6 & 0.150 \\
Run$\_{40}\_1.5\_3g$ & 40 & 1.5 & 767 & 1.0 & 30 & 2 & 30.6 & 44.0 & 0.0025 & 11.3 & 0.196 \\
Run$\_{40}\_1.5\_4g$ & 40 & 1.5 & 767 & 1.0 & 30 & 4 & 28.9 & 24.0 & 0.0121 & 6.2 & 0.341 \\
Run$\_{40}\_1.5\_5g$ & 40 & 1.5 & 767 & 1.0 & 30 & 5 & 26.4 & 18.5 & 0.0891 & 7.4 & 0.313 \\
Run$\_{40}\_1.5\_6g$ & 40 & 1.5 & 767 & 1.0 & 30 & 5 & 13.9 & 20.2 & 0.0057 & 7.3 & 0.280 \\
Run$\_{40}\_1.5\_7g$ & 40 & 1.5 & 767 & 1.0 & 30 & 4 & 19.2 & 22.8 & 0.0097 & 6.7 & 0.317 \\
Run$\_{40}\_1.5\_8g$ & 40 & 1.5 & 767 & 1.0 & 30 & 3 & 18.7 & 35.5 & 0.0225 & 6.9 & 0.242 \\
Run$\_{40}\_1.5\_9g$ & 40 & 1.5 & 767 & 1.0 & 30 & 5  & 23.3 & 18.3 & 0.0063 & 5.6 & 0.350 \\
Run$\_{40}\_1.5\_10g$ & 40 & 1.5 & 767 & 1.0 & 30 & 3 & 32.9 & 34.1 & 0.0277 & 8.0 & 0.241 \\
Run$\_{40}\_1.5\_11g$ & 40 & 1.5 & 767 & 1.0 & 30 & 11 & 7.2 & 15.4 & 0.0007 & 6.1 & 0.442 \\
Run$\_{40}\_1.5\_12g$ & 40 & 1.5 & 767 & 1.0 & 30 & 1 & 43.2 & $\cdots$ & 0.163 & $\infty$ & 0.106 \\
Run$\_{50}\_1\_1g$ & 50 & 1.0 & 767 & 1.5 & 30 & 4 & 28.2 & 20.9 & 0.0083 & 7.9 & 0.511 \\
Run$\_{50}\_1\_2g$ & 50 & 1.0 & 767 & 1.5 & 30 & 3 & 37.8 & 28.8 & 0.0011 & 10.6 & 0.344 \\
Run$\_{50}\_1\_3g$ & 50 & 1.0 & 767 & 1.5 & 30 & 3 & 51.1 & 27.9  & 0.0243 & 8.3 & 0.549 \\
Run$\_{50}\_1\_4g$ & 50 & 1.0 & 767 & 1.5 & 30 & 4 & 39.8 & 21.2 & 0.0042 & 8.6 & 0.511 \\
Run$\_{50}\_1\_5g$ & 50 & 1.0 & 767 & 1.5 & 30 & 3 & 55.4 & 25.9  & 0.0151 & 13.8 & 0.491 \\
Run$\_{80}\_1\_1g$ & 80 & 1.0 & 767 & 1.5 & 22 & 2 & 45.2 & 28.4 & 0.101 & 21.9 & 0.368 \\
Run$\_{80}\_1\_2g$ & 80 & 1.0 & 767 & 1.5 & 22 & 4 & 57.3 & 18.1 & 0.0078 & 9.1 & 0.580 \\
Run$\_{80}\_1\_3g$ & 80 & 1.0 & 767 & 1.5 & 22 & 3 & 49.5 & 25.7 & 0.0290 & 10.2 & 0.400 \\
Run$\_{80}\_1\_4g$ & 80 & 1.0 & 767 & 1.5 & 22 & 3 & 50.9 & 26.6 & 0.0352 & 10.6 & 0.339 \\
Run$\_{80}\_1\_5g$ & 80 & 1.0 & 767 & 1.5 & 22 & 2 & 57.3 & 31.2 & 0.0041 & 14.3 & 0.433 \\
Run$\_{80}\_1\_6g$ & 80 & 1.0 & 767 & 1.5 & 22 & 3 & 48.9 & 25.8 & 0.0086 & 7.4 & 0.500 \\
\hline
Run$\_{50}\_1.5\_1N$ & 50 & 1.5 & $\cdots$ & $\cdots$ & 25 & 3 & 11.8 & 33.3 & 0.0200 & 9.4 & 0.215 \\
Run$\_{50}\_1.5\_2N$ & 50 & 1.5 & $\cdots$ & $\cdots$ & 25 & 3 & 10.6 & 30.9 & 0.0305 & 10.4 & 0.194 \\
Run$\_{50}\_1.5\_3N$ & 50 & 1.5 & $\cdots$ & $\cdots$ & 25 & 3 & 14.5 & 30.8 & 0.0173 & 8.2 & 0.264 \\

\enddata
\end{deluxetable}

\clearpage

\begin{deluxetable}{lccccc|cccccc}
\tablecolumns{12}
\tablewidth{0pc}
\tablecaption{Simulation Inputs and Outputs. A simulation named Demo$\_{X}\_{Y}\_{n}$ is the nth realisation of a 
disk of mass X $M_{\oplus}$, with a surface density power law index $\alpha=Y$.
\label{Democheck}}
\tablehead{
 &
\multicolumn{4}{c}{Inputs}& & &\multicolumn{5}{c}{Outputs} \\
\colhead{Name} & \colhead{$M_Z$}   & \colhead{$\alpha$}  & \colhead{$M_{gas}$} & \colhead{$\beta$} & \colhead{$N_{tot}$} &
\colhead{N($<1.1 AU$)} &  \colhead{$M_{big} $} & \colhead{$S_s$} & \colhead{AMD} & \colhead{$S_{c}$} & \colhead{$<a>_M$} \\
  & \colhead{($M_{\oplus}$)} & & \colhead{($M_{\oplus}$)} & & & & \colhead{($M_{\oplus}$)} & & & & \colhead{(AU)} }
\startdata
Demo$\_{50}\_1.5\_1$ & 50 & 1.5 & $\cdots$ & $\cdots$ & 52 & 3 & 23.7 & 31.1 & 0.0254 & 6.5 & 0.379 \\
Demo$\_{50}\_1.5\_2$ & 50 & 1.5 & $\cdots$ & $\cdots$ & 52 & 3 & 15.0 & 28.4 & 0.0214 & 13.1 & 0.291 \\
                     & 50 & 1.5 &  $\cdots$ & $\cdots$ & 52 & 4 & 15.0 & 25.3 & 0.0199 & 9.7 & 0.356 \\
Demo$\_{50}\_1.5\_3$ & 50 & 1.5 &  $\cdots$ & $\cdots$ & 52 & 2 & 17.0 & 45.3 & 0.0467 & 13.2 & 0.169 \\
                     & 50 & 1.5 &  $\cdots$ & $\cdots$ & 52 & 3 & 17.0 & 33.4 & 0.0471 & 7.6 & 0.272 \\
Demo$\_{50}\_1.5_4$  & 50 & 1.5 &  $\cdots$ & $\cdots$ & 52 & 2 & 22.0 & 45.0 & 0.0495 & 13.9 & 0.174 \\
                     & 50 & 1.5 &  $\cdots$ & $\cdots$ & 52 & 3 & 22.0 & 36.0 & 0.0498 & 5.0 & 0.422 \\
Demo$\_{50}\_1.5_5$  & 50 & 1.5 &  $\cdots$ & $\cdots$ & 52 & 3 & 21.0 & 34.2 & 0.0218 & 8.4 & 0.291 
\enddata
\end{deluxetable}

\clearpage

%\plotone{f1.ps}

%\figcaption[f1.ps]{
%Planet mass versus orbital period. The shaded areas show the approximate loci of planet classes observed in the Howard et al. (2010) paper. 
%In each panel, the solid circles, open circles and crosses represent disk masses (interior to 1 AU) of
%100 $M_{\oplus}$, 50 $M_{\oplus}$ and 25 $M_{\oplus}$ respectively (except for the uppermost panel, where
%the crosses represent the results of the Q=1.5 simulations). The lower panel represents the results of
%simulations with a surface density profile with $\alpha=0$. The middle panel represents the results of
%simulations with a surface density profile with $\alpha=3/2$ and the upper panel uses $\alpha=5/2$. For
%each simulation, the planets shown are those remaining (summed over 3 realisations) after 10~Myr. The
%shaded region is best matched with $\alpha=3/2$ and disk masses between 50 and 100 $M_{\oplus}$.
%\label{Ma1}}

\plotone{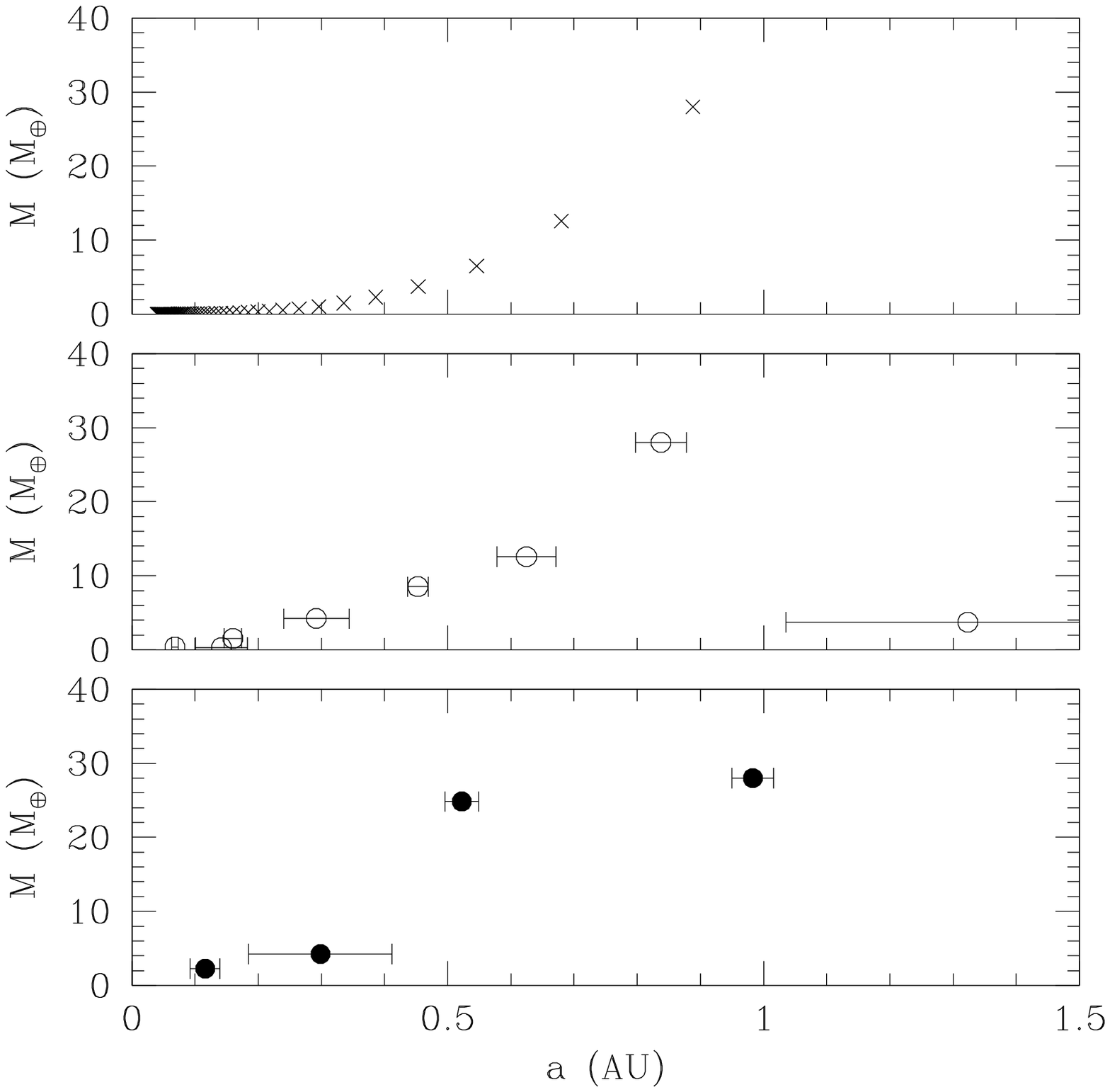}

\figcaption[f1.ps]{The three panels in this figure show the mass and semi-major axis distribution of a 
population of bodies realised (Run$\_50\_0\_7$) from a constant surface density profile, at three different ages. The total mass
in all bodies is $50 M_{\oplus}$, initially spread between 0.05 and 1~AU. The top 
panel shows the initial conditions, with the mass distributed amongst bodies according to the oligarchic model
of Kokubo \& Ida (1998). The middle panel shows the population after 0.1~Myr of dynamical evolution. The horizontal
error bars indicate the radial excursions due to finite eccentricity. We see that the inner 0.2~AU is already
significantly evolved at this time. The lower panel shows the same population after 10~Myr of evolution. The mass
is now distributed amongst four surviving bodies, with the mass budget dominated by the outer pair.
\label{Snap0}}

\clearpage

\plotone{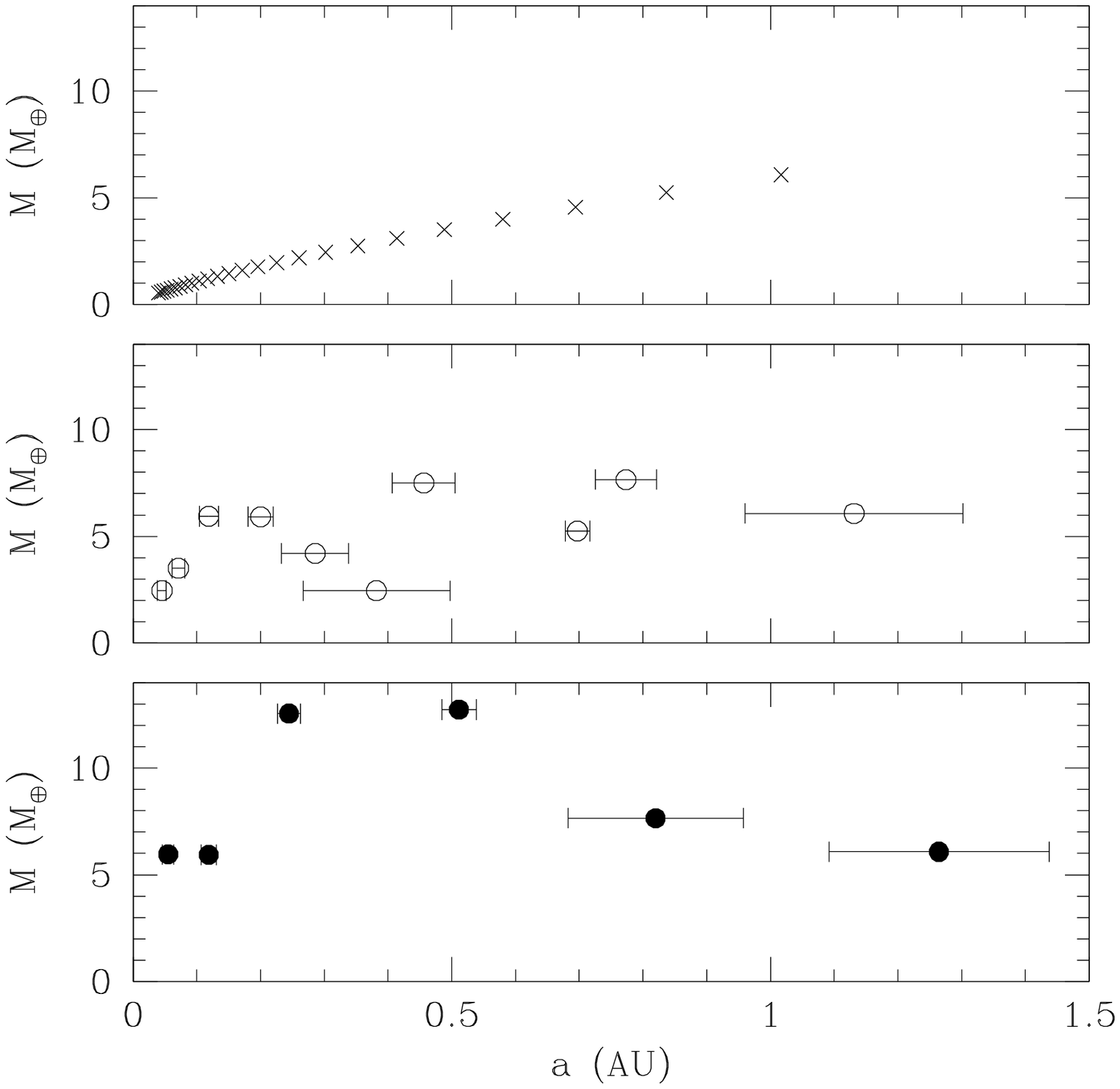}

\figcaption[f2.ps]{The three panels in this figure show the mass and semi-major axis distribution of a 
population of bodies realised (Run$\_50\_1.5\_10$) from a surface density profile $\propto a^{-1.5}$, at three different ages. The total mass 
in all bodies is $50 M_{\oplus}$, initially spread between 0.05 and 1~AU. The top panel shows the initial conditions, with the mass distributed amongst bodies according to the oligarchic model
of Kokubo \& Ida (1998). The middle panel shows the population after 0.1~Myr of dynamical evolution. The horizontal
error bars indicate the radial excursions due to finite eccentricity. In this model, the faster dynamical evolution and
accumulation at small scales leads to an approximately constant distribution of mass as a function of semi-major axis.
 The lower panel shows the same population after 10~Myr of evolution. The mass is once again distributed amongst a handful of surviving bodies, 
but with a more even distribution.
\label{Snap1.5}}

\clearpage

\plotone{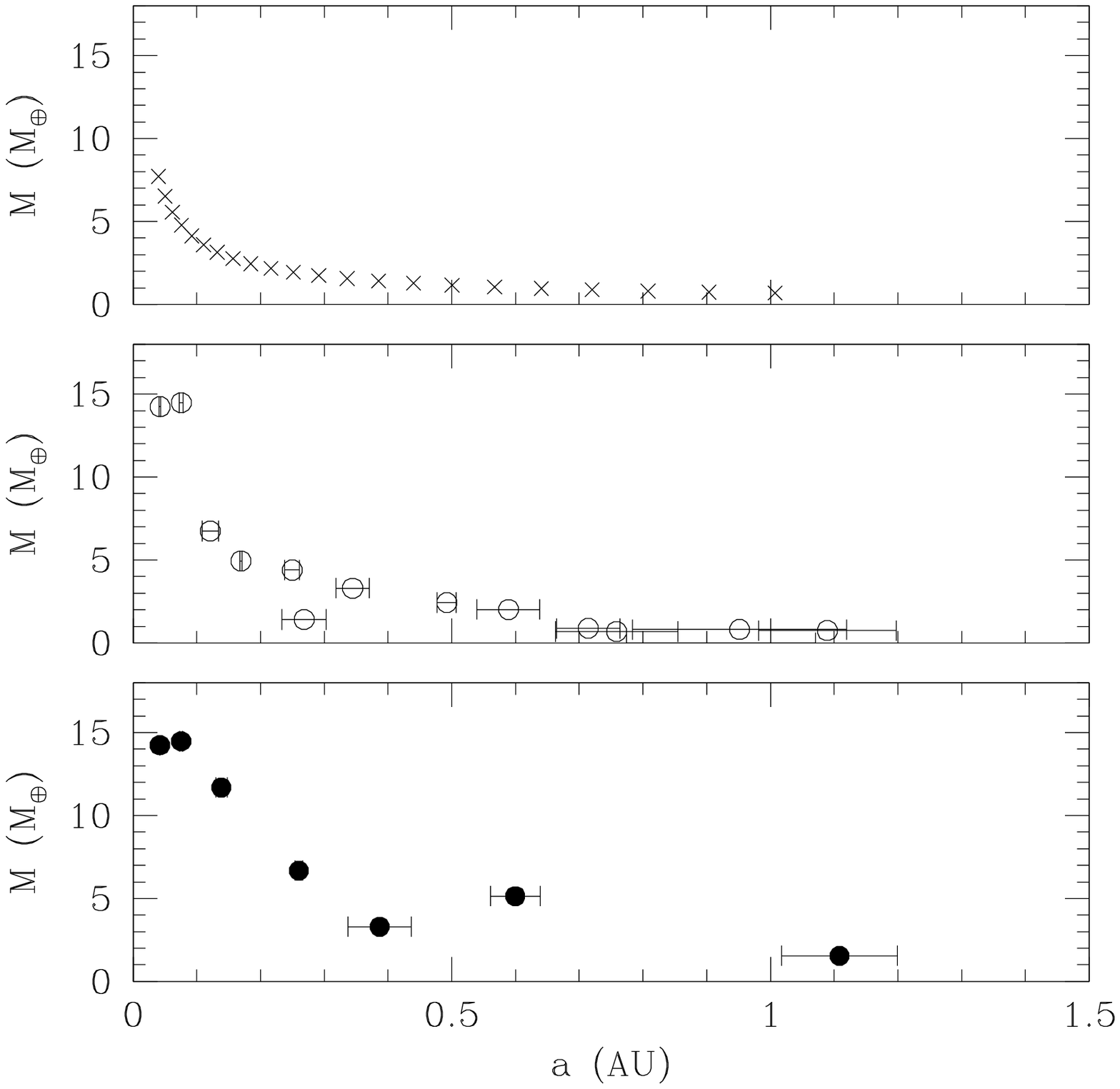}

\figcaption[f3.ps]{The three panels in this figure show the mass and semi-major axis distribution of a
population of bodies realised (Run$\_50\_2.5\_8$) from a surface density profile $\propto a^{-2.5}$, at three different ages. The total mass
in all bodies is $50 M_{\oplus}$, initially spread between 0.05 and 1~AU. The top panel shows the initial conditions, with the mass distributed amongst bodies according to the oligarchic model
of Kokubo \& Ida (1998). The middle panel shows the population after 0.1~Myr of dynamical evolution. The horizontal
error bars indicate the radial excursions due to finite eccentricity. In this model, the faster dynamical evolution and
accumulation at small scales leads to an approximately constant distribution of mass as a function of semi-major axis.
 The lower panel shows the same population after 10~Myr of evolution. The steepness of the mass distribution means that
the bulk of the mass remains collected amongst a few bodies on small scales. Scattering processes do not lead to significant
outward mass transport.
\label{Snap2.5}}

\clearpage

\plotone{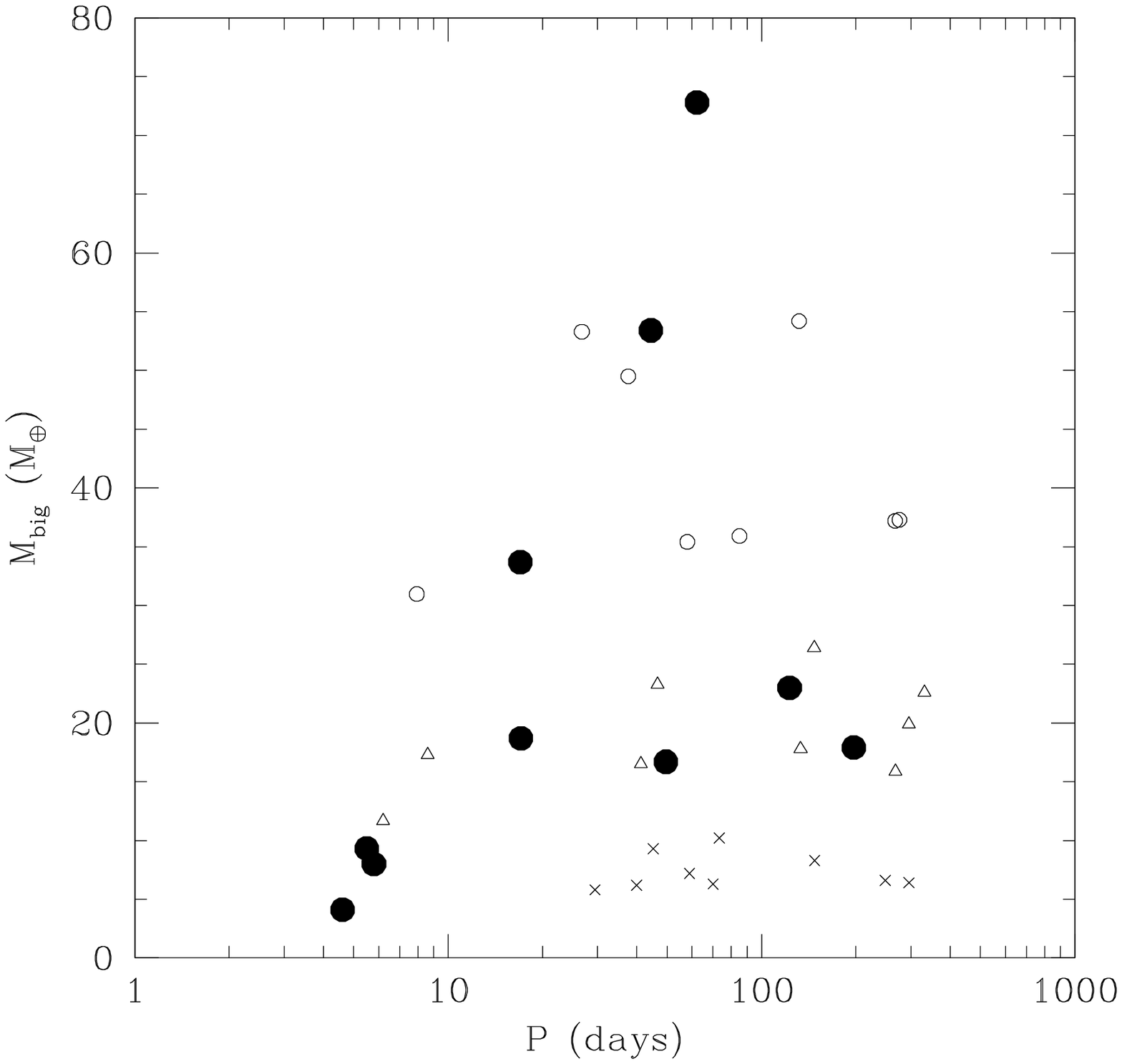}

\figcaption[f4.ps]{The solid circles indicate the orbital periods and masses of the largest bodies in each system with an
observed planet in the sample of Howard et al (Jupiter-mass or longer period planets are to be found off the edges of this plot). The
open circles are the corresponding values for the largest surviving body in a series of simulations with a disk of total mass
$100 M_{\oplus}$ and power law slope $\alpha=1.5$. The open triangles are the corresponding values for simulations with the
same slope and total mass of $50 M_{\oplus}$. The crosses are for a total disk mass of 25 $M_{\oplus}$.
\label{Param2_4}}

\clearpage

\plotone{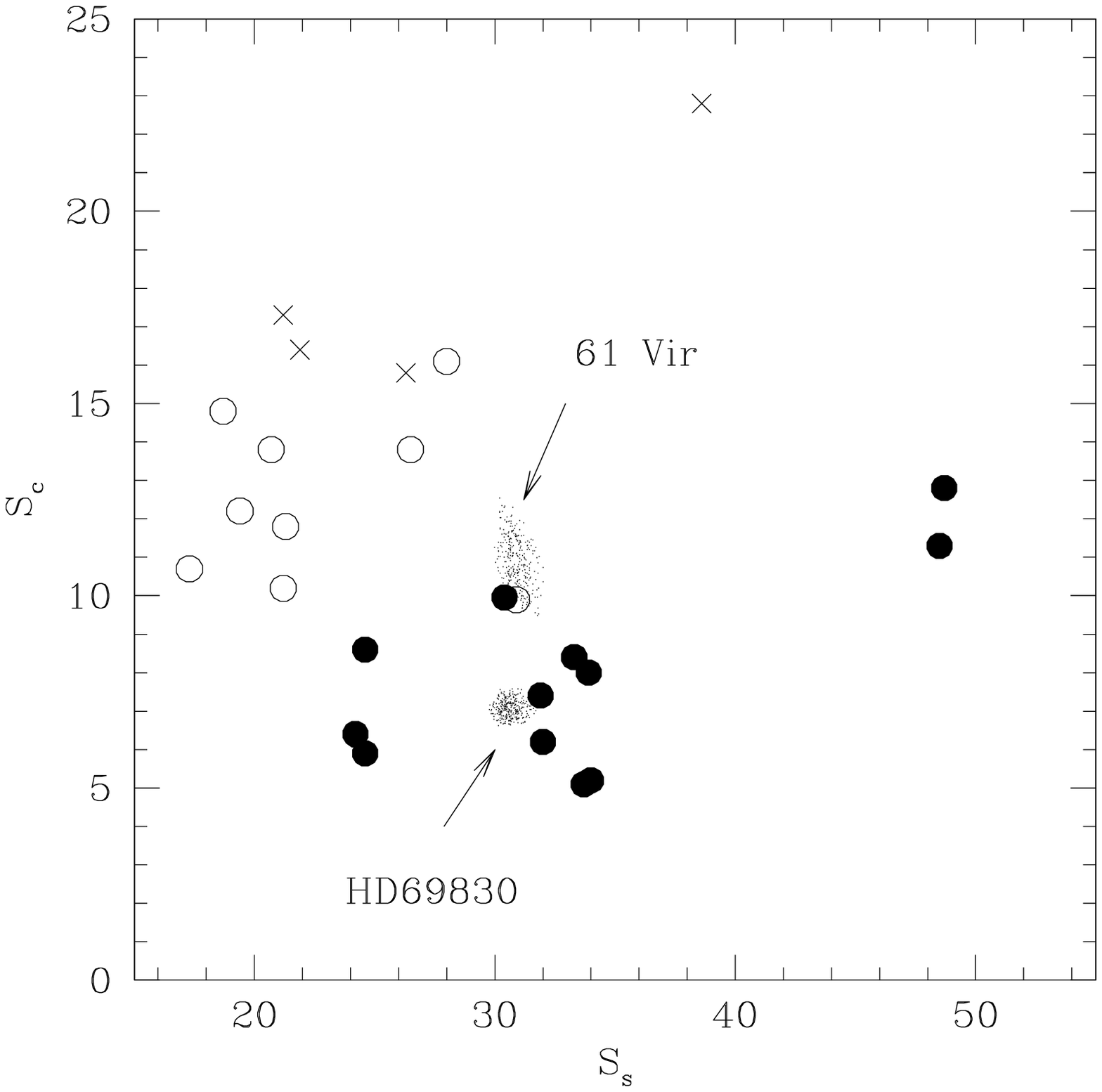}

\figcaption[f5.ps]{
Mass concentration $S_c$ versus average orbital spacing $S_s$ (both quantities are defined in appendix~\ref{Stats}). The open circles, filled circles, and crosses show the statistical measures for 9 simulations each for disk profiles
$\alpha=5/2$, $\alpha=3/2$ and $\alpha=0$ respectively. In each case, the total disk mass interior to 1~AU was $50 M_{\oplus}$.
 The clouds of small dots indicate the parameter range for the 61~Virginis
 and HD69830 systems, allowing for uncertainties in the system parameters. We see that these observed systems exhibit a degree of mass concentration best 
matched by the $\alpha=3/2$ models.  Only four crosses are shown because many of the $\alpha=0$ runs produce $S_c>25$.
\label{Param2}}

\clearpage

\plotone{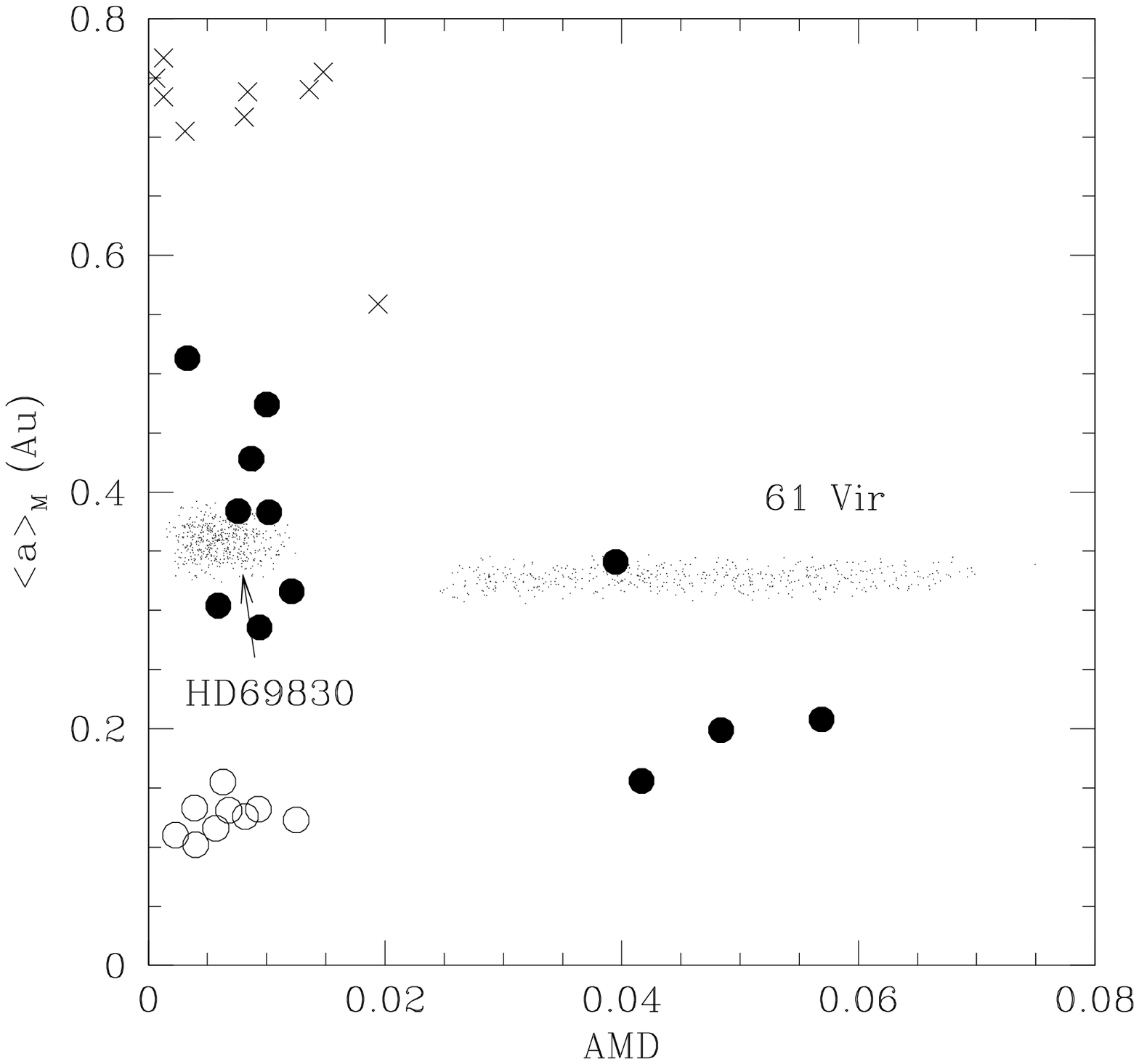}

\figcaption[f6.ps]{The mass-weighted average semi-major axis $\langle a\rangle_M$ versus the angular momentum deficit $AMD$ (both defined in appendix~\ref{Stats}). The open circles, filled circles, and crosses show $\langle a\rangle_M$ and $AMD$ for 3 simulations each for disk profiles
$\alpha=5/2$, $\alpha=3/2$ and $\alpha=0$ respectively. In each case, the total disk mass interior to 1~AU was $50 M_{\oplus}$.
The regions of small dots indicate the parameter range for the 61~Virginis
 and HD69830 systems, allowing for uncertainties in the system parameters. Once again, the $\alpha=3/2$ models fit the observations
better. Flatter(Steeper) density profiles concentrate the mass on scales that are too large(small) compared to the observed systems. The filled circles with $\langle a \rangle_M<0.3$AU are systems in which one of the surviving planets does not produce a large enough ($>1 m/s$) radial velocity
signature, and so is not counted.
\label{Param3}}

%\plotone{f4.ps}

%\figcaption[f4.ps]{The left hand panel shows the same parameter space as Figure~\ref{Param2}, and the right hand panel
%reproduces Figure~\ref{Param3}. However, the solid points now represent an extended set of realisations of only one scenario -- the model with
%$\alpha=3/2$ and 50 $M_{\oplus}$ of rocks interior to 1~AU. We see that both the 61 Vir and HD69830 systems are consistent
%with being drawn from this model.
%\label{Param5}}

\clearpage
\plotone{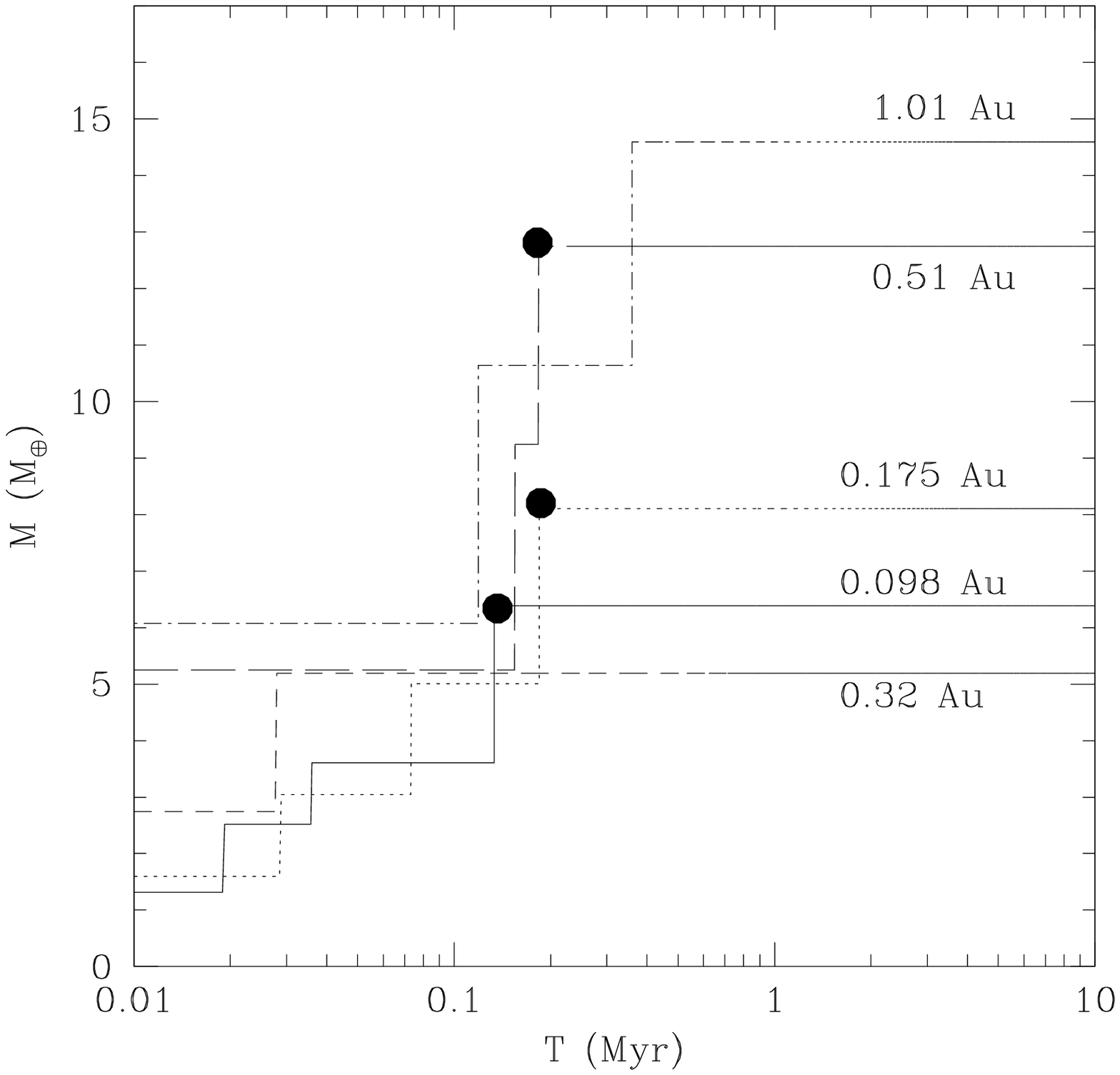}
\figcaption[f7.ps]{The five curves show the accumulation history in one particular system (Run$\_{50}\_{1.5}\_12$), starting with
$50 M_{\oplus}$. Each curve traces the mass evolution of a planet that survives to the end, and is labelled with the final location of the planet.
The simulation started with 25 bodies. We see that the mass of
each final planet is largely accumulated within 1~Myr. The three filled circles show when planets cross the
threshold defined by equation~(\ref{GapMaker}). Note that the lowest mass surviving planet does not contribute to the tabulated statistics,
as it produces a radial velocity signature $< 1 m/s$. \label{Acc}}

\clearpage

\plotone{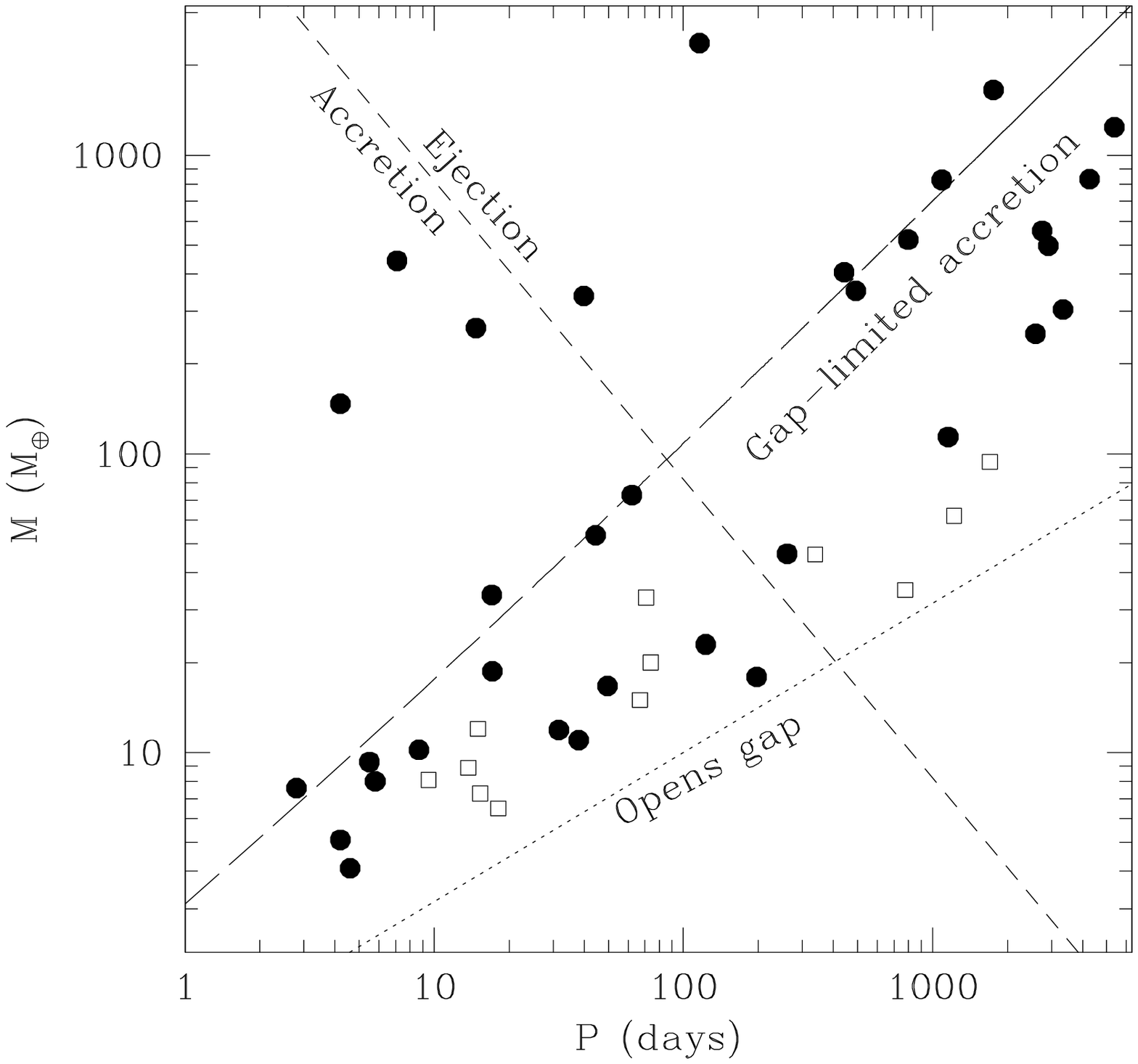}
\figcaption[f8.ps]{The mass-period diagram. The solid and open points are the planetary detections and candidates, respectively, announced by Howard et al. (2010). 
The short dashed line running from upper left to lower right is the boundary between accretion and ejection for a body accreting smaller objects through 
gravitational encounters. The dotted line indicates the mass required to accrete gas and thereby open a gap in the natal gas disk, according to equation (\ref{GapMaker}). The long-dashed curve (from eqn. \ref{GapLimit}) shows the planet mass that results if a core,
 mass determined by the dotted line, accretes the gas required to open a gap in the disk, assuming a gas disk mass interior to 1~AU of four times the MMSN.
%The dot-dashed diagonal line at the lower left indicates the approximate threshold below which gaseous atmospheres may be photoevaporated over the course of the
%system lifetime.
\label{PM}}

\clearpage

\plotone{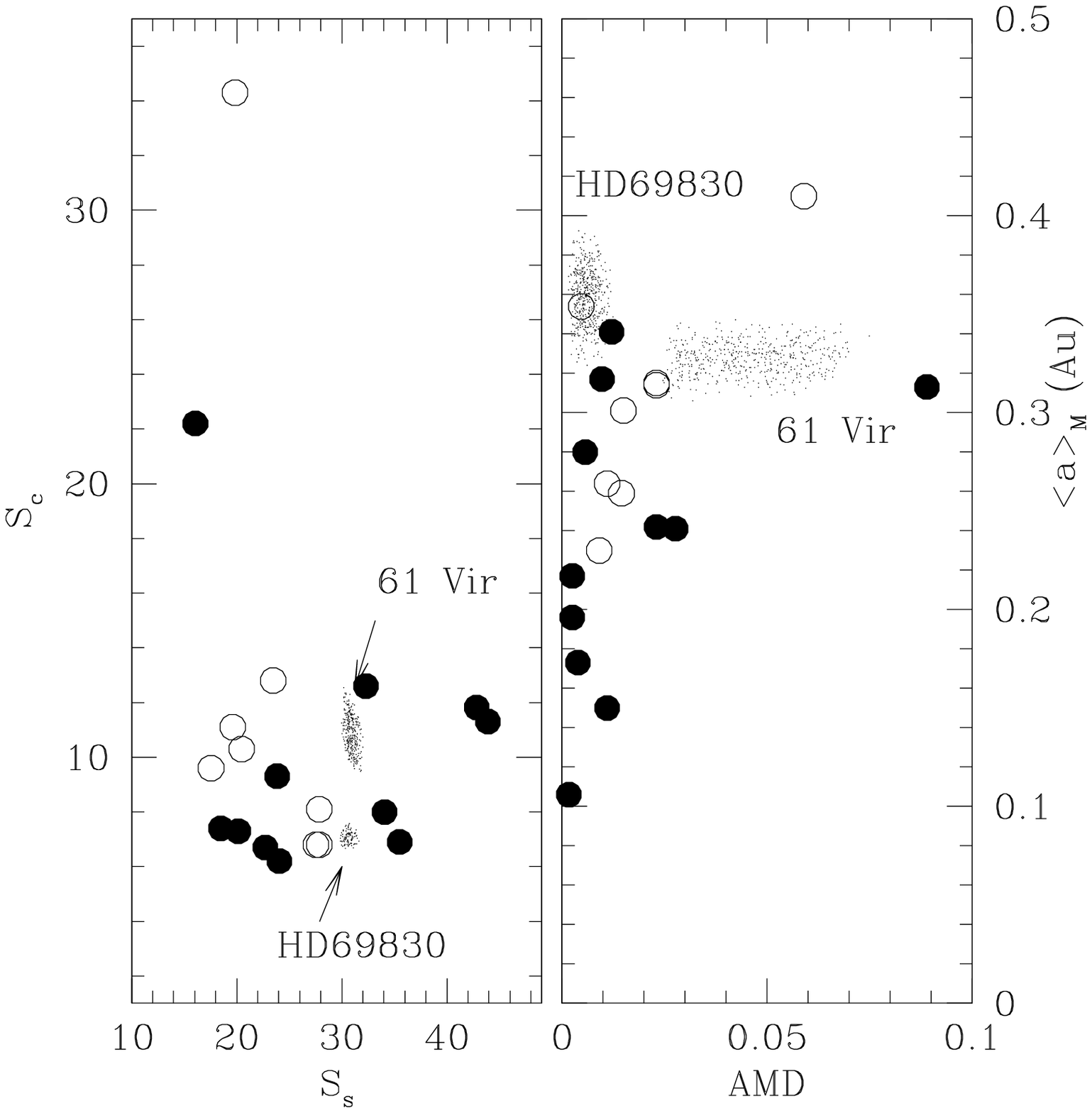}
\figcaption[f9.ps]{The panels are the same as in Figures~\ref{Param2} and \ref{Param3}, except that the points now show the models in which gas was accreted when planets
crossed the threshold in equation~(\ref{GapMaker}). The open circles are for the more massive planet model ($50 M_{\oplus}$ rock disk, Gas disk $3\times$MMSN)
and the filled circles are for the less massive model ($40 M_{\oplus}$ rock disk, MMSN Gas disk).
\label{Param4}}

\clearpage

\plotone{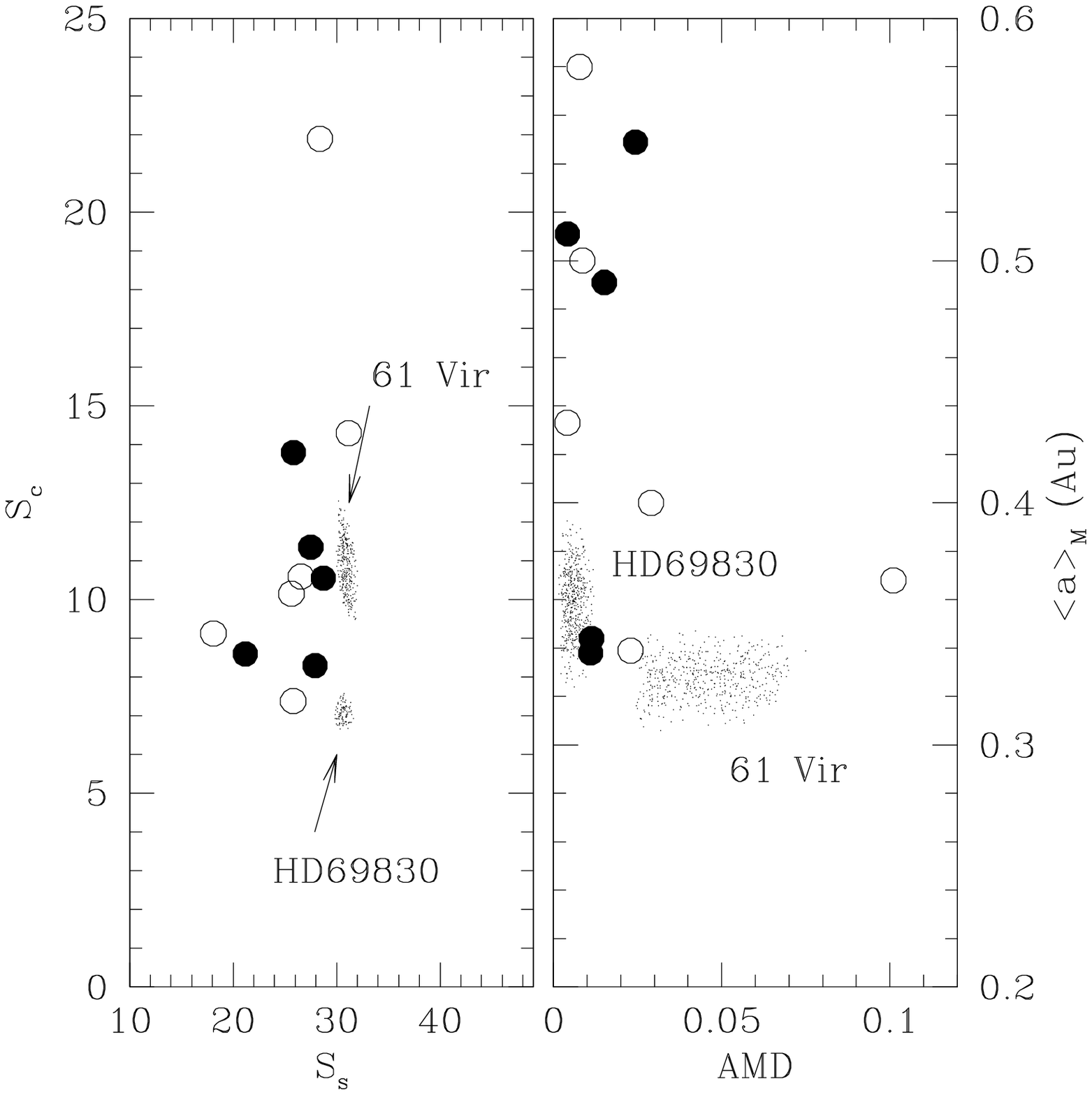}
\figcaption[f10.ps]{The panels are the same as in Figure~\ref{Param4}, except that the rocky cores are drawn from $\alpha=1$ profiles. Solid points
indicate a disk normalisation of 50$M_{\oplus}$ and open points are for 80~$M_{\oplus}$.
\label{Param7}}
\clearpage

\plotone{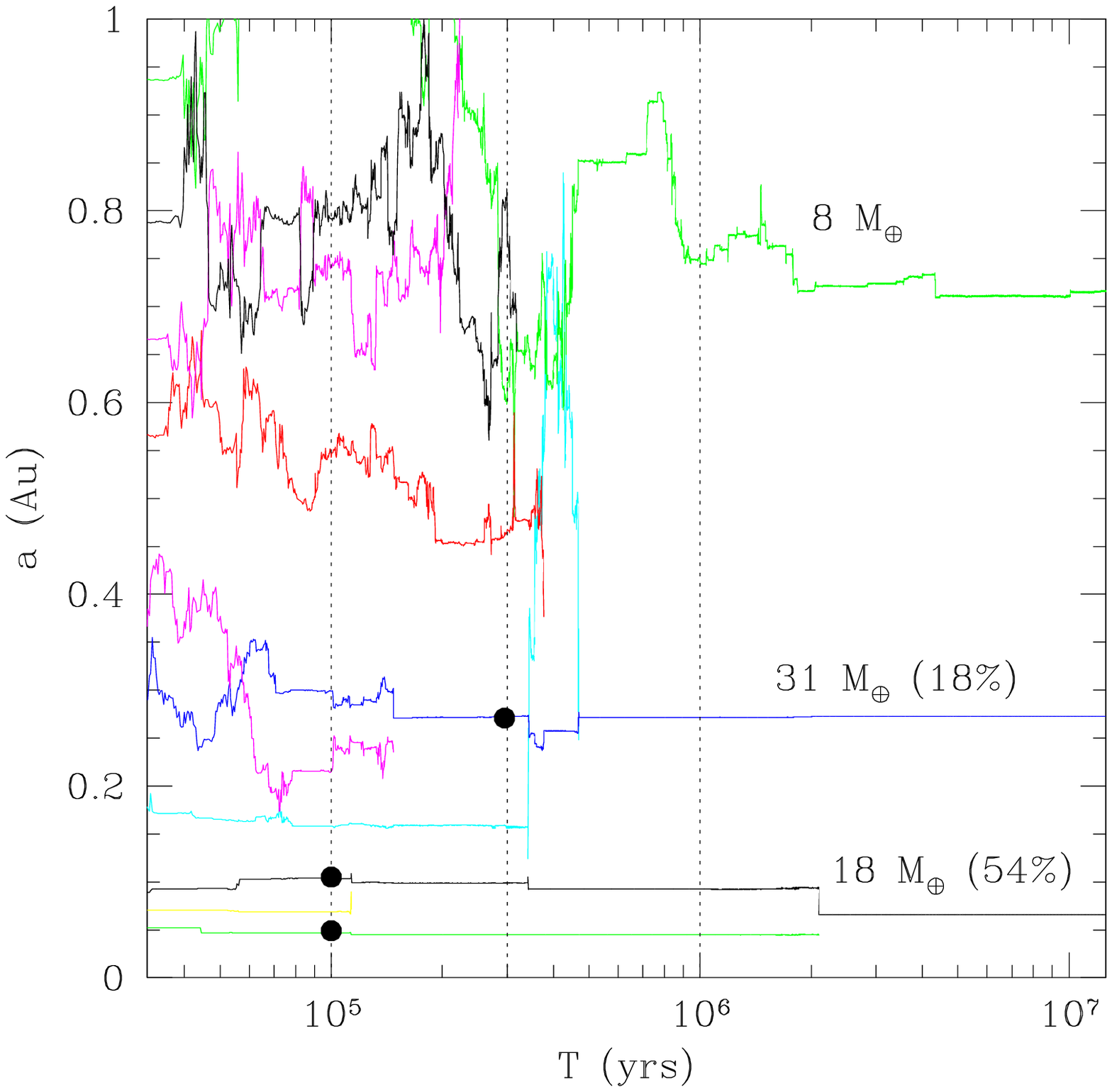}
\figcaption[f11.ps]{Semimajor axis versus time for planets that survive for at least one hundred thousand years in a simulation (Run$\_40\_1.5\_3g$) that features tidally-limited gas accretion. The vertical dashed lines indicate where we stopped the simulation and estimated whether any objects had crossed the
threshold (\ref{GapMaker}). The three solid points indicate which planets accreted gas and at which time. Curves that end abruptly indicate that the
body was lost by accretion onto another object. The final masses of the three surviving bodies are indicated, along with the gas mass fractions
for the two inner bodies that accreted gas from the disk. The late time evolution of the outer body is driven by scattering and eventual ejection
of another body, off the top of the plot, on an eccentric orbit.
\label{TA22}}

\clearpage

\plotone{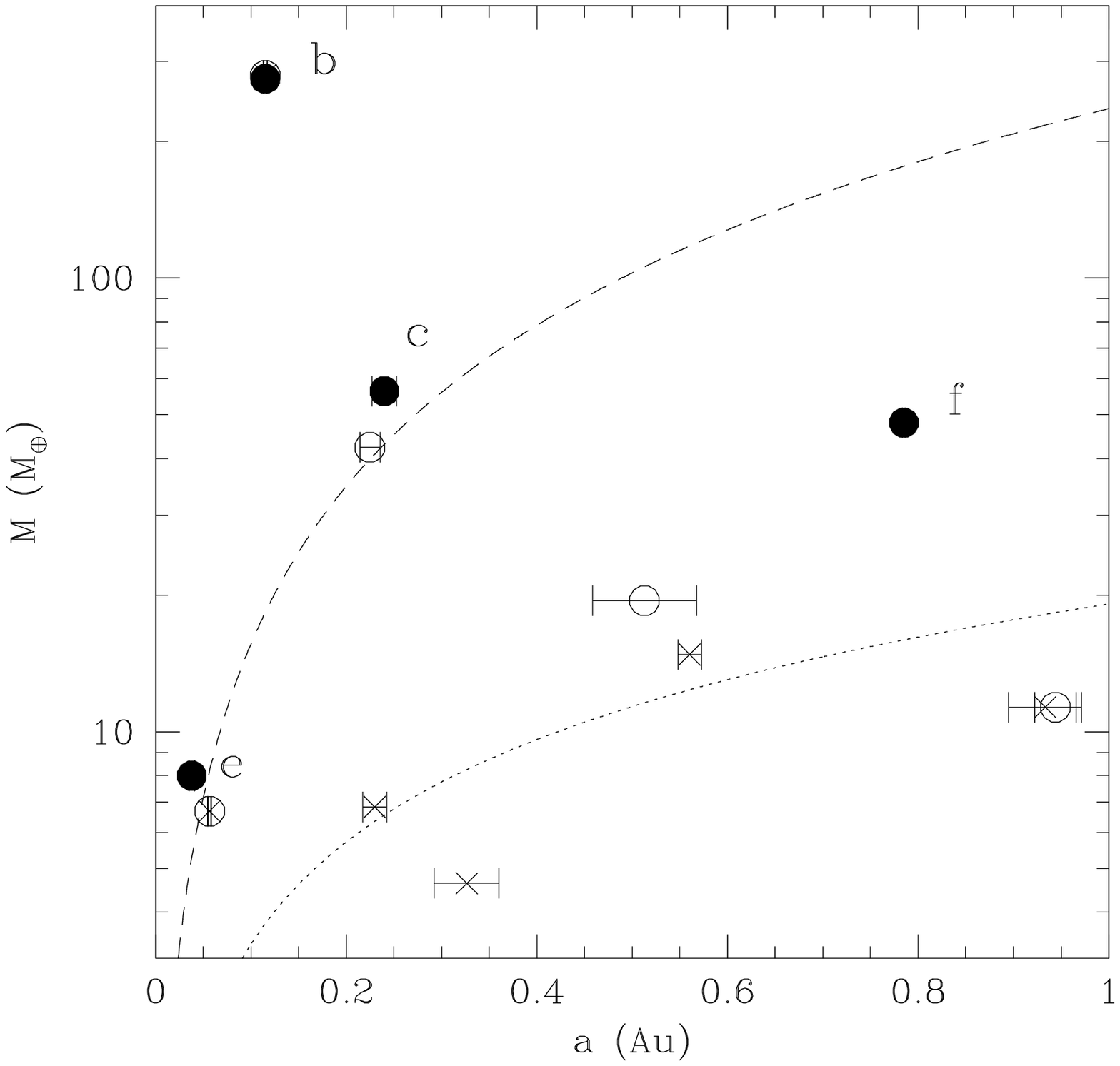}
\figcaption[f12.ps]{The labelled solid points are the inferred minimum masses for the observed 55~Cancri system (there is one additional known planet not shown
because it lies significantly to the right on this scale). 
 The crosses show the result of an in situ accumulation of rocky planets after 5~Myr. The open circles show the system wherein the mass of the planet
near 0.2~AU was enhanced artificially to simulate limited gas accretion, and then run for an additional 5~Myr. The dotted line indicates our estimated threshold mass
for gas accretion and the dashed line the estimated mass after gas accretion. It is encouraging that our scenario can produce the appropriate configuration of planets
close to the star -- we did not attempt to exactly match analogoues of planet~f as the separation is large enough that we do not expect the perturbations from
planet~b to have a dynamically significant effect. The parameters of the innermost planet, 55~Cancri~e, have been recently revised (Dawson \& Fabrycky 2010; Winn et al. 2011b)
such that the planet lies even closer to the star than shown here. We have verified, by running a shorter simulation of the inner disk, extended down to 0.02~AU, that the
presence of 55~Cancri~b does not prevent the formation of planets with orbital periods $<$~1~day.
\label{55match}}

\clearpage

\plotone{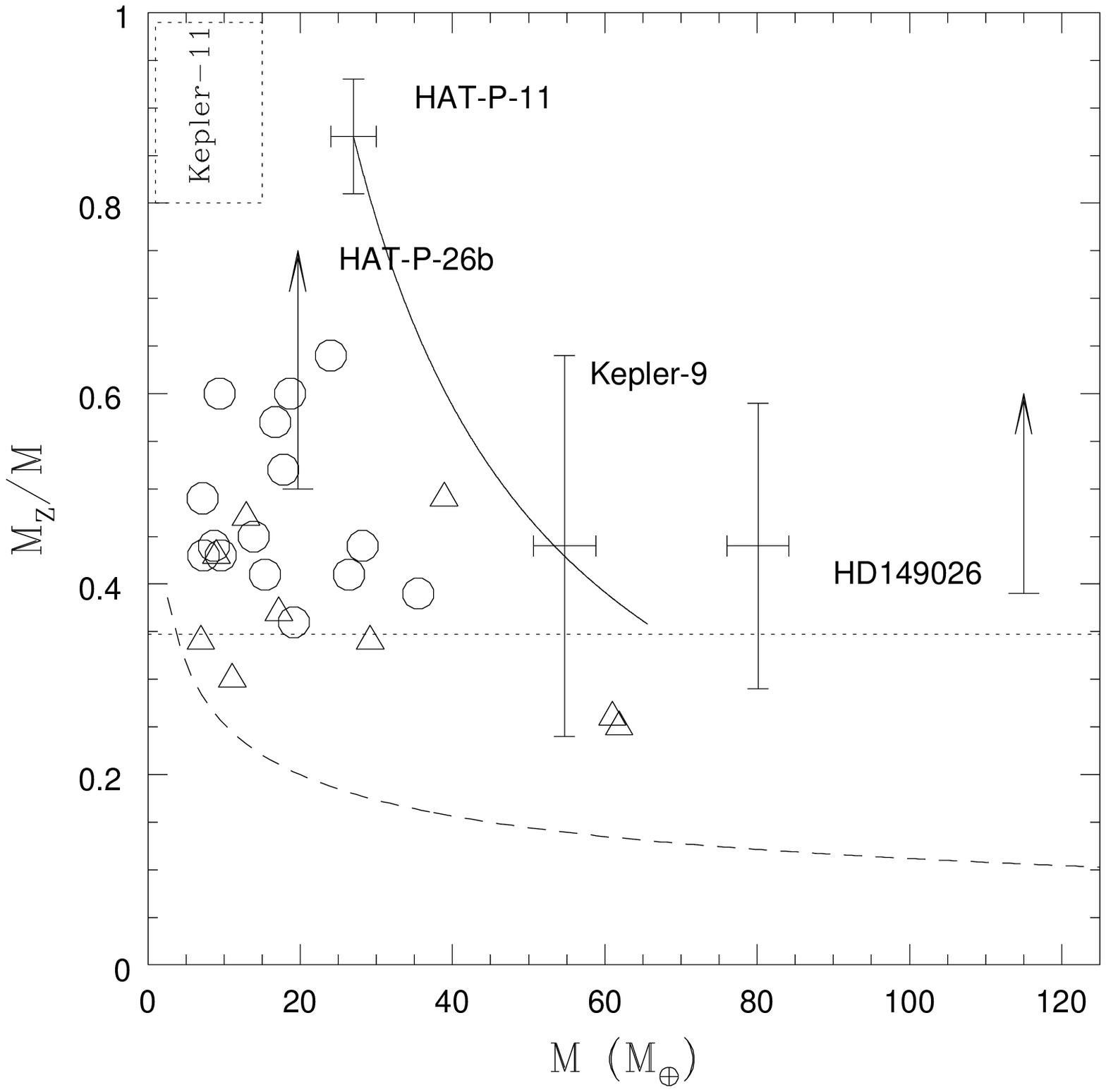}
\figcaption[f13.ps]{Planet metallicity versus planet mass. The open circles and triangles represent planets realised in the simulations described in the text. The dotted and dashed lines correspond
to the models for gas accretion described by equations~(\ref{GasScale}) and (\ref{GapLimit}) as realised for the same simulations. The fact that the final planets
lie above these curves is the result of accretion of extra solids during the clearing process. The box in the upper left indicates the range of estimated parameters
for the planets in the Kepler-11 system. At the extreme upper left ($M_Z/M=1$ and $M<10 M_{\oplus}$) is where one would find the
rocky cores of evaporated planets as realised in our simulations (assuming the envelopes of  planets with orbital periods inside 10 days are evaporated). The errorbars and arrows indicate the published
values for the labelled planets from Burrows et al. (2007), Carter et al. (2009), Hartman et al. (2009), Rogers \& Seager (2010) and Havel et al. (2011).
The solid line extending from the HAT-P-11 point indicates the path traced if we add metal-free gas to the planet (to simulate the replacement of losses to evaporation).
\label{MX}}

\clearpage

\plotone{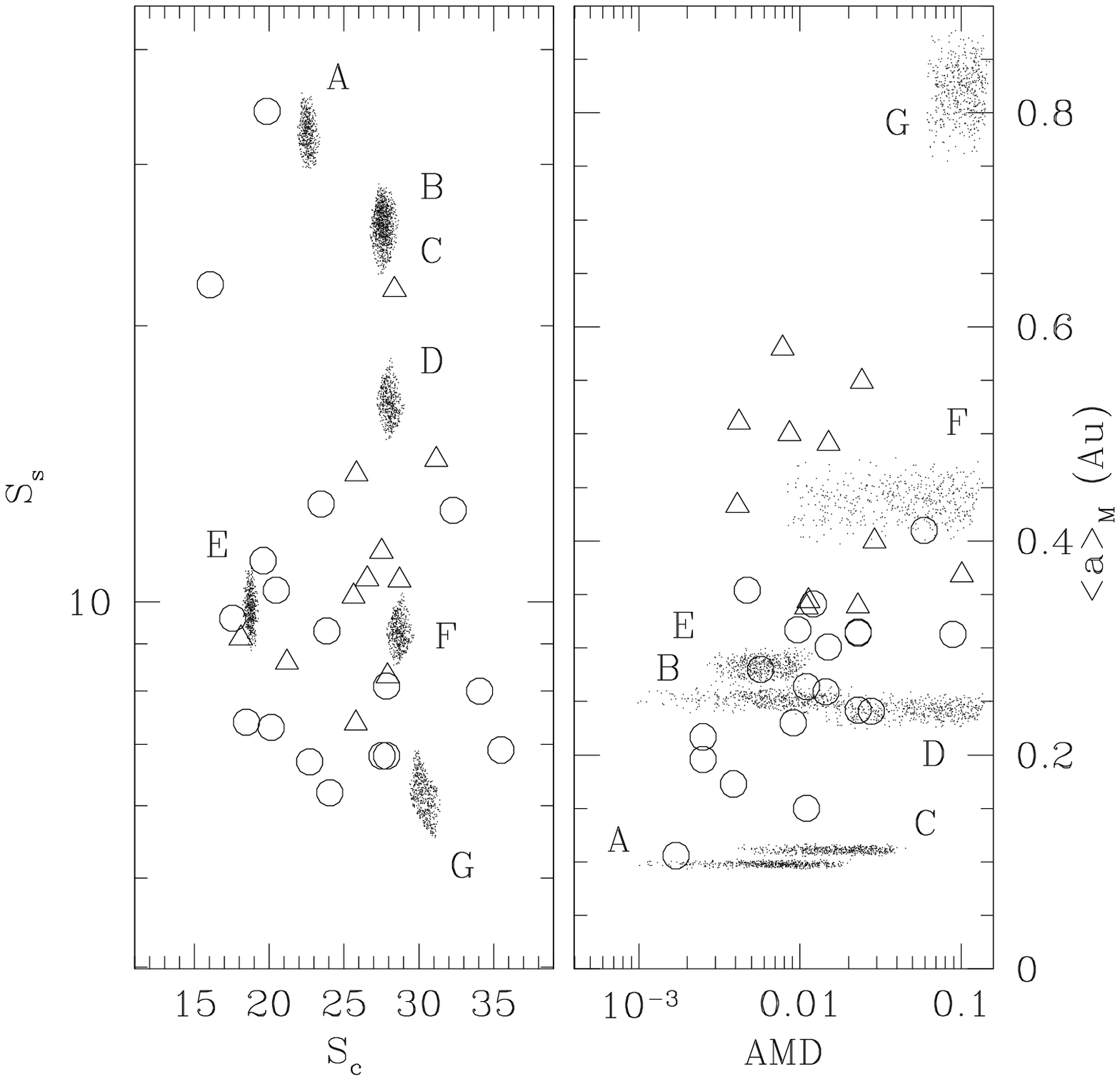}
\figcaption[f14.ps]{The left and right hand panels show the parameters for seven known three-planet planetary systems  in the sample of Mayor et al. (2011), all of which have three or more
announced sub-Jupiter planets inside 1.2~AU. Each system's region of parameter space is marked with the corresponding letter. The systems are 
HD~40307 (A), HD~20794 (B), HD~39194 (C), HD~136352 (D), HD~10180 (E), HD~31527 (F) and HD~134606 (G).
 The open symbols show the results from the simulations shown in Figure~\ref{Param4} (circles) and 
Figure~\ref{Param7} (triangles). In cases where eccentricities were not quoted, we assumed a range from 0 to 0.1.
\label{Param6}}

\clearpage

\plotone{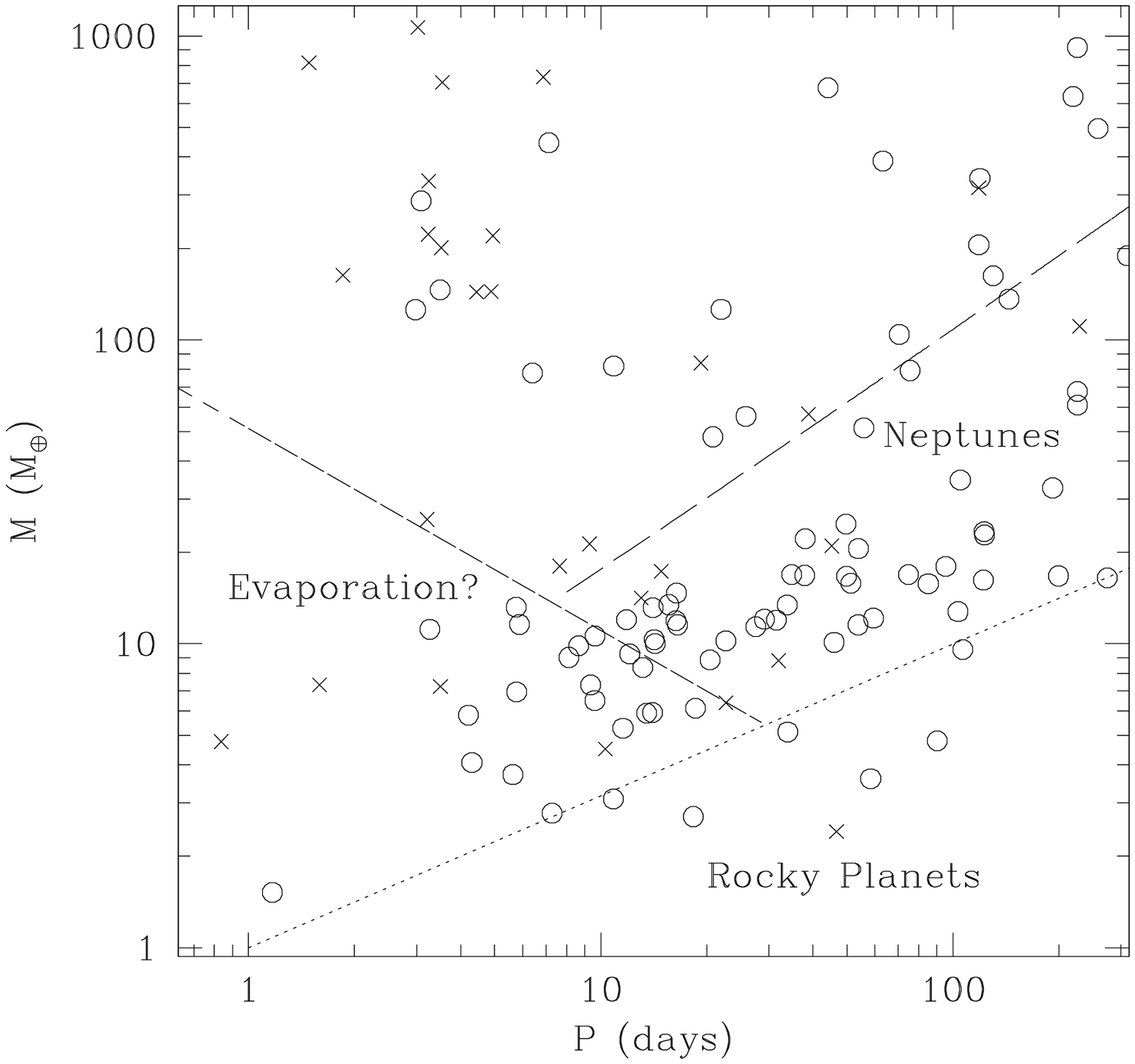}
\figcaption[f15.ps]{Mass-Period diagram for two other planet samples. The open points are the planets detected in the Mayor et al. (2011) sample studied with the HARPS and Coralie instruments. The crosses indicate planets drawn from the Kepler transit sample and confirmed by radial velocities.
The lines delineate 
parts of parameter space that, according to our model, are occupied by genuine Rocky planets (below the dotted line), Neptunes (rocky cores with comparable mass gaseous envelopes, between the long-dashed and dotted lines), and
(partially) evaporated Neptunes (above the dotted line, down and to the left from the short-dashed line, given by eqn. \ref{EvapLim} with $\eta=1$).
\label{pmall}}


\begin{references}
\reference{AB} Adams, F. C. \& Bloch, A. M., 2009, ApJ, 701, 1381
\reference{Ag} Agnor, C. B., Canup, R. M. \& Levison, H. F., 1999, Icarus 142, 219
\reference{Ali} Alibert, Y. et al., 2006, A\&A, 455, L25
\reference{AL} Artymowicz, P. \& Lubow, S. H., 1996, ApJ, 467, L77
\reference{BS10} Bai, X.-N. \& Stone, J. M., 2010, ApJ, 722, 1437
\reference{Bat} Batalha, M., et al., 2011, ApJ, 729, 27
\reference{BHL} Bodenheimer, P., Hubickyj, O. \& Lissauer, J. J., 2000, Icarus, 143, 2
%\reference{Bon} Bonfils, X., 2005, A\&A, 443, L15
\reference{Kep1} Borucki, W. B. et al., 2011, ApJ, 736, 19
\reference{BCL} Bryden, G., Chen, X., Lin, D. N. C., Nelson, R. P. \& Papaloizou, J. C. B., 1999, ApJ, 514, 344
\reference{Bu07} Burrows, A. et al., 2007, ApJ, 661, 502
\reference{CW} Carter, J. A., Winn, J. N, Gilliland, R. \& Holman, M. J., 2009, ApJ, 696, 241
\reference{C99} Chambers, J. E., 1999, MNRAS, 304, 793
\reference{C01} Chambers, J. E., 2001, Icarus, 152, 205
\reference{CW} Chambers, J. E. \& Wetherill, G. W., 1998, Icarus, 136, 304
\reference{CWB} Chambers, J. E., Wetherill, G. W. \& Boss, A. P., 1996, Icarus, 119, 261
\reference{DF} Dawson, R. I. \& Fabrycky, D. C., 2010, ApJ, 722, 937
\reference{ED} Ehrenreich, D., \& D\'{e}sert, J.-M., 2011, arXiv:1103.0011
\reference{FV} Fischer, D. A. \& Valenti, J., 2005, ApJ, 622, 1102
\reference{G01} Gammie, C., 2001, ApJ, 553, 174
\reference{GT} Goldreich, P. \& Tremaine, S., 1980, ApJ, 241, 425
\reference{G98} Gonzalez, G., 1998, A\&A, 334, 221
\reference{Han} Hansen, B. M. S., 2009, ApJ, 703, 1131
\reference{H11}  Hartman, J. D., et al., 2011, ApJ, 728, 138
\reference{Hav} Havel, M., Guillot, T., Valencia, D. \& Crida, A., 2011, arXiV:1103.6020
\reference{Hay} Hayashi, C., 1981, Prog. Theor. Phys. Suppl., 70, 35
\reference{Hol} Holman, M. et al., 2010, Science, 330, 51
\reference{How} Howard, A. W. et al., 2010, Science, 330, 653
\reference{How11} Howard, A. W. et al., 2011, ApJ, 726, 73
\reference{IdaLin} Ida, S. \& Lin, D. N. C., 2008, ApJ, 685, 84
\reference{ID10} Ida, S. \& Lin, D. N. C., 2010, ApJ, 719, 810
\reference{IEN} Ikoma, M., Emori, H. \& Nakazawa, K, 2001, ApJ, 553, 999
\reference{K99} Kley, W., 1999, MNRAS, 303, 696
\reference{KN} Kley, W. \& Nelson, R. P., 2012, arXiv:1203.1184
\reference{KKI} Kokubo, E., Kominami, J. \& Ida, S., 2006, ApJ, 642, 1131
\reference{KI98} Kokubo, E. \& Ida, S., 1998, Icarus, 131, 171
\reference{KI02} Kokubo, E. \& Ida, S., 2002, ApJ, 581, 666
\reference{LSR} Lammer, H., Selsis, F., Ribas, I., Guinan, E., Bauer, S. J., \& Weiss, W., 2003, ApJ, 598, L121
\reference{LKA} Laughlin, G., \& Rozyczka, M., 1996, ApJ, 456, 279 
\reference{Leg} Leger, A., et al., 2009, A\&A, 506, 287
\reference{LcE04} Lecavelier des Etangs, A., Vidal-Madjar, A., McConnell, J. C. \& Hebrard, G., 2004, ApJ, 621, 1049
\reference{LcE} Lecavelier des Etangs, A., 2007, A\&A, 461, 1185
\reference{LBR} Lin, D. N. C., Bodenheimer, P. \& Richardson, D. C., 1996, Nature, 380, 606
\reference{LP} Lin, D. N. C. \& Papaloizou, J., 1979, MNRAS, 186, 799
\reference{LFF} Lissauer, J. J., et al., 2011b, Nature, 470, 53
\reference{Kep2} Lissauer, J. J. et al., 2011a, ApJ, 197, 8
\reference{Lov} Lovis, C. et al., 2006, Nature, 441, 305
\reference{Lov2} Lovis, C. et al., 2011, A\&A, 528, A112
\reference{MPR} Matsumura, S., Peale, S. J. \& Rasio, F. A., 2010, ApJ, 725, 1995
%\reference{MBF} Marcy, G., Butler, P., Vogt, S., Fischer, D. \& Lissauer, J., 1998, ApJ, 505, L147
\reference{MBF2} Marcy, G., Butler, P., Fischer, D., Vogt, S., Lissuaer, J. \& Rivera, E., 2001, ApJ, 556, 296
\reference{MQ} Mayor, M. \& Queloz, D., 1995, Nature, 378, 355
\reference{Maya} Mayor, M. et al., 2009a, A\&A, 493, 639
\reference{Mayb} Mayor, M. et al., 2009b, A\&A, 507, 487
\reference{May11} Mayor, M. et al., 2011, arXiv:1109.2497
\reference{McN} McNeil, D. S. \& Nelson, R. P., 2010, MNRAS, 401, 1691
\reference{MA} Moeckel, N. \& Armitage, P. J., 2012, MNRAS, 419, 366
\reference{Mord} Mordasini, C., Alibert, Y. \& Benz, W., 2009, A\&A, 501, 1139
\reference{M98} Murray, N., B. Hansen, M. Holman \& S. Tremaine, 1998, Science, 279, 69
\reference{McIM} Murray-Clay, R. A., Chiang, E. I. \& Murray, N., 2009, ApJ, 693, 23
\reference{NSH} Nakagawa, Y., Sekiya, M. \& Hayashi, C., 1986, Icarus, 67, 375
\reference{OMM} Oishi, J. S., Mac~Low, M.-M. \& Menou, K., 2007, ApJ, 670, 805
\reference{PM} Paardekooper, S. J. \& Mellema, G., 2006, A\&A, 459, L17
\reference{Poll96} Pollack, J. B., Hubickyj, O., Bodenheimer, P., Lissauer, J. J., Podolak, M. \& Greenzweig, Y., 1996, Icarus, 124, 62
\reference{Raf06} Rafikov, R., 2006, ApJ, 648, 666
\reference{RF96} Rasio, F. A. \& Ford, E. B., 1996, Science, 274, 954
\reference{RBM} Raymond, S. N., Barnes, R. \& Mandell, A., 2008, MNRAS, 384, 663
\reference{RQL} Raymond, S. N., Quinn, T. \& Lunine, J. I., 2004, Icarus, 168, 1
\reference{RQL2} Raymond, S. N., Quinn, T. \& Lunine, J. I., 2005, ApJ, 632, 670
\reference{RA} Rice, W. K. M.  \& Armitage, P. J., 2009, MNRAS, 396, 2228
%\reference{RLB} Rivera, E., et al., 2005, ApJ, 634, 625
\reference{RS} Rogers, L. A. \& Seager, S., 2010, ApJ, 712, 974
\reference{SIM} Santos, N. C., Israelian, G. \& Mayor, M., 2001, A\&A, 373, 1019
\reference{SS} Shakura, N. I. \& Sunyaev, R. A., 1973, A\&A, 24, 337
\reference{TI} Taniguchi, T. \& Ikoma, M., 2007, ApJ, 667, 55
\reference{TP} Terquem, C. \& Papaloizou, J. C. B., 2007, ApJ, 654, 1110
\reference{TTP} Tian, F., Toon, O. B., Pavlov, A. A. \& de Sterck, H., 2005, ApJ, 621, 1049
\reference{US} Udry, S. \& Santos, N. C., 2007, ARA\&A, 45, 397
%\reference{UBD} Udry, S. et al., 2007, A\&A, 469, 43
\reference{Vog} Vogt, S. et al., 2009, ApJ, 708, 1366
%\reference{VBR} Vogt, S., Butler, P., Rivera, E., Haghighipour, N., Henry, G. \& Williamson, M., 2010, ApJ, 723, 954
\reference{W97} Ward, W. R., 1997, Icarus, 126, 261
\reference{W77a} Weidenschilling, S. J., 1977a, Ap\&SS, 51, 153
\reference{W77b} Weidenschilling, S. J., 1977b, MNRAS, 180, 57
\reference{Winn} Winn, J.N., et al., 2011a, AJ, 141, 63
\reference{Winn} Winn, J. N. et al., 2011b, ApJ, 737, L18
\reference{WL11} Wolfgang, A. \& Laughlin, G., 2011, arXiV:1108.5842
\reference{Yelle} Yelle, R. V., 2004, Icarus, 170, 167
\reference{YS02} Youdin, A. \& Shu, F., 2002, ApJ, 580, 494
\reference{You11} Youdin, A., 2011, ApJ, 742, 38
\end{references}
\end{document}